\documentclass[preprint,showpacs,preprintnumbers,amsmath,amssymb,nofootinbib]{revtex4-1}

\usepackage{graphicx}
\usepackage{dcolumn}
\usepackage{bm}

\newcommand{\beq}{\begin{equation}}
\newcommand{\eeq}{\end{equation}}
\newcommand{\beqy}{\begin{eqnarray}}
\newcommand{\eeqy}{\end{eqnarray}}
\newcommand{\beqyn}{\begin{eqnarray*}}
\newcommand{\eeqyn}{\end{eqnarray*}}
\newcommand{\nl}{\newline}
\newcommand{\nn}{\nonumber}

\newcommand{\slas}[1]{\not\!{#1}}

\newcommand{\bc}{\begin{center}}
\newcommand{\ec}{\end{center}}
\newcommand{\bmin}{\begin{minipage}}
\newcommand{\emin}{\end{minipage}}

\begin{document}

\title{On the controversy concerning the definition of quark and gluon angular momentum}

\author{Elliot Leader}
 \email{e.leader@imperial.ac.uk}
\affiliation{Blackett laboratory \\Imperial College London \\ Prince Consort Road\\ London SW7 2AZ, UK}

\date{\today}

\begin{abstract}
 A major controversy has arisen in QCD as to how to split the total angular momentum  into separate quark and gluon contributions,  and as to whether the gluon angular momentum can itself be split, in a gauge-invariant way, into a spin and orbital part. Several authors have proposed various answers to these questions and offered a variety of different expressions for the relevant  operators. I argue that none of these is acceptable and suggest that the \emph{canonical}  expression for the momentum and angular momentum operators  is the correct and physically meaningful one. It is then an inescapable fact that the  gluon  angular momentum operator cannot, in general, be split in a gauge-invariant way into a spin and orbital part. However,  the projection of the gluon spin onto its direction of motion i.e. its helicity is gauge invariant and is measured in deep inelastic scattering on nucleons. The Ji sum rule, relating the quark angular momentum to generalized parton distributions, though not based on the canonical operators, is shown to be correct, if interpreted with due care. \nl
 I also draw attention to several interesting aspects of QED and QCD, which, to the best of my knowledge, are not commented upon in the standard textbooks on Field Theory.

\end{abstract}

\maketitle

\section{\label{Intro} Introduction}
A major controversy has arisen in QCD as to how to split the total angular momentum into separate  quark and gluon components (throughout this paper ``quark" will mean a sum over all flavours of quarks \emph{and} antiquarks). The idea of identifying separate quark and gluon angular momentum operators is attractive, since these operators may be measurable in certain physical processes and there may be sum rules relating the spin of a nucleon to the angular momentum carried by its constituents. The operators for total momentum and total angular momentum, obtained via Noether's theorem from the QCD Lagarangian, consist of separate terms which seem to represent a natural division into quark and gluon pieces. However,  Ji, in particular, \cite{Ji:1997pf, *Ji:1996ek} has argued that such terms are not individually gauge invariant and has advocated use of the Bellinfante version of these operators,  which   has the nice property that they are gauge invariant and can be measured in Deeply-virtual Compton Scattering reactions \cite{Ji:1996nm}. But Ji's quark angular momentum operator contains both quark fields  \emph{and}  the gluon vector potential, so is not obviously to be interpreted as the physical quark angular momentum. Indeed, a  major debate has arisen as to whether it is correct to identify this operator as the quark angular momentum, and Chen, Lu, Sun, Wang and Goldman \cite{Chen:2008ag} and Wakamatsu \cite{Wakamatsu:2010qj} have proposed  quite different identifications, leading to very different statements as to what fractions of momentum and angular momentum the quarks and gluons carry in the asymptotic limit $Q^2\rightarrow \infty $. In Ji's Bellinfante approach no attempt is made to split the gluon angular momentum into a spin part and an orbital part, in accord with the long held belief that such a splitting cannot be done in a gauge-invariant way. But both  Chen et al \cite{Chen:2008ag} and Wakamtsu \cite{Wakamatsu:2010qj} claim much more, namely, that it is possible to carry our such a division in a gauge-invariant way and that even in QED the traditional, decades-old textbook method of identifying electron and photon angular momentum is incorrect! ( For access to the papers in the controversy see ref.\cite{Chen:2009dg}.) \nl
The paper of Wakamtsu \cite{Wakamatsu:2010qj} explains very clearly how the differences between the various approaches arise. In QED one splits the photon vector potential into two parts

\beq \label{Asplit} \bm{A} = \bm{A}_{phys} + \bm{A}_{pure} \eeq
corresponding exactly to what is usually called the transverse $\bm{A}_\bot $ and longitudinal $\bm{A}_\|$ parts respectively, with
\beq \label{Aphys} \bm{\nabla}\centerdot \bm{A}_{phys}=0 \quad \textrm{and} \quad \bm{\nabla} \times \bm{A}_{pure}=0 \eeq
Under a gauge transformation $\bm{A}_{phys}$ is invariant, whereas
\beq \label{Apure} \bm{A}_{pure}(x) \rightarrow \bm{A}_{pure}(x) + \bm{\nabla}\Lambda(x) \eeq
In QCD, analogously, one splits
\beq A^\mu_a = A^\mu_{phys,a} + A^\mu_{pure,a} \eeq
where $A^\mu_{pure,a}$ transforms like $A^\mu_a $ itself under gauge transformations, but is a pure gauge in the sense that it gives rise to no non-zero fields i.e. $G^{\mu\nu}_{pure} = 0 $, while $A^\mu_{phys,a}$ transforms covariantly i.e. like $G^{\mu\nu}$ itself. \nl
Wakamatsu shows that the difference between the various versions lies in the freedom to insert a particular term
\beq \bm{V}\equiv g\int d^3x \psi^\dag_l(x) (\bm{x}\times\bm{A}^a_{phys})t^a_{lm}\psi_m(x) \eeq
either into the quark orbital angular momentum or into the gluon angular momentum, yielding, he claims, two possibilities. But, in fact, if there is no other criterion to indicate which is the correct choice, there is actually an infinite number of possibilities i.e. one could insert $\alpha \bm{V}$ into the quark orbital term and $(1-\alpha)\bm{V}$ into the gluon term. \nl
In a later paper \cite{Wakamatsu:2010cb} Wakamatsu attempted to reformulate his approach in a manifestly covariant form and to relate his spin and orbital terms to the polarized parton densities which are measured in polarized deep inelastic scattering. Unfortunately many of the equations in this paper are incorrect as a result of treating a \emph{non-forward} matrix element like \nl $\langle \,p + \Delta/2; S \,| \, M^{\mu\nu\lambda}  \, | \,p-\Delta /2; S \,  \rangle $ as transforming like a tensor, and forgetting that the physical requirement on the covariant polarization vector, namely $S\cdot (p\pm \Delta /2) =0$ implies $S\cdot \Delta =0$. It should be stressed that the existence of these errors is not controversial. The same errors occur in the Jaffe-Manohar paper \cite{Jaffe:1989jz} and have been graciously acknowledged by those authors\footnote{private communication from Professor Jaffe to T. L.Trueman and the author}. Umfortunately then, it is very difficult to decide which claims in the Wakamatsu paper are justified.

We shall argue that none of these prescriptions is generally correct or physically plausible, but we shall see that the Bellinfante version works in certain specific situations. There are three main problems:

1) In all these papers much emphasis is placed on the issue of using gauge-invariant operators. We shall show that this emphasis is misplaced and that the gauge invariance of the \emph{operators} is not an important criterion. In particular we suggest that neither Ji's, Chen et al's nor Wakamtsu's identification is  physically correct. We shall first show below that in any theory which is invariant under gauge transformations, even the   \emph{total} momentum and angular momentum operators \emph{cannot} be gauge invariant. Of course this does \emph{not} mean that the momentum and angular momentum cannot be measured. Because what one measures---and this is the key point--- are not operators but matrix elements of operators, and if care is exercised in defining the physical states of the theory (respecting any subsidiary conditions, which is crucial in a gauge theory) then these matrix elements turn out to be gauge invariant. This is the basis for our  suggestion that the emphasis on utilizing gauge invariant \emph{operators} is misleading. Then we shall discuss what happens \emph{if} one insists on using  gauge invariant operators and demonstrate that they  do  not, in general, have the physical meaning expected of them.

2) In all the above papers the treatment is essentially classical and use is made of the classical equations of motion. This totally ignores the highly non-trivial complications involved in quantizing a gauge theory and the fact that some classical equations cannot be maintained at the operator level. For example in QED,  when one writes for the photon vector potential the symbol $A_\mu (x)$,  it creates the expectation that it transforms like a 4-vector under Lorentz transformations. Yet to agree with the Maxwell equations $A_\mu (x)$  has to satisfy a subsidiary condition. In classical electrodynamics one chooses the beautiful covariant Lorenz condition $\partial^\mu  A_\mu (x) = 0 $, which indeed permits $A_\mu (x)$  to transform as a 4-vector. It is well known, however, that one cannot impose such a subsidiary condition on the \emph{operators} $A_\mu (x)$ in QED, since it contradicts the usual canonical equal-time commutation relations of the quantized theory. There are many  approaches to the quantization of electrodynamics in which a non-covariant subsidiary condition is imposed (for a concise summary  see Section 21.2 of \cite{Leader:1996hm}). A  popular choice is the Coulomb gauge condition
$\bm{\nabla}\centerdot \bm{A} = 0 $  (see, for example, Section 13.5 of \cite{Bjorken:1965dk}) . If this gauge condition is to hold in any reference frame then clearly   $A_\mu (x)$ cannot behave as a 4-vector, but---and this is the crucial point---this does \emph{not}  spoil the Lorentz invariance of the theory, since the \emph{matrix elements} corresponding to any measurable physical quantity do transform correctly \cite{Zumino:60}. Thus, firstly,  we suggest that it is unnecessary to insist that $A_\mu (x)$ transforms as a 4-vector, and secondly, but more importantly,  if, as Ji does,  one does insist that one's vector potential is a genuine 4-vector, then one has to deal with a \emph{covariantly} quantized theory, in which case the expressions given in  the Ji,  Chen et al and Wakamatsu papers, for the linear and angular momenta, are incomplete. The covariant quantization of QED is a non-trivial task \cite{Lautrup, Nakanishi:66,Nakanishi:72} involving the introduction of a scalar gauge-fixing field $B(x)$. Covariant QCD is even more complicated, both in instant form  \cite{Kugo:78} and light-front form \cite{Srivastava:1999gi}, involving both a gauge-fixing field and Faddeev-Popov ghosts fields. In both QED and QCD  the expressions for the linear and angular momentum should include terms involving all these fields.

3) The key issue of splitting the total momentum and angular momentum into a  quark and gluon contribution is not adequately analyzed. There are two rather separate aspects. There is the age-old question of splitting the angular momentum of a gauge particle into a spin part and an orbital part. We shall discuss this in Section~\ref{helicity}. But there is a more general question of how, in any theory with interacting fields, say $\phi_E(x)$ and $\phi_F(x)$, one can split the total momentum (and angular momentum) into pieces interpretable as the contributions of the quanta $E$ and $F$. In all the above papers, having invented some strategy for defining the operators $\bm{P}_E$ and $\bm{P}_F$,  one writes, for the total momentum
\beq \label{splitP} \bm{P}= \bm{P}_E + \bm{P}_F \eeq
and then interprets the nucleon expectation values of theses operators as a measure of the contribution of $E$ and $F$ respectively to the momentum of the nucleon. But this is potentially  misleading, because the interacting particles constantly exchange momentum, and the correct way to express Eq.~(\ref{splitP}) is
\beq \label{splitPgood} \bm{P}= \bm{P}_E(t) + \bm{P}_F(t) \eeq
to reflect the fact that while the total momentum is conserved, the individual momenta are not. Thus it requires some analysis to explain why it is meaningful to interpret, e.g.
$ \langle  \textrm{nucleon}\,|\, \bm{P}_E(t) \, |\, \textrm{nucleon} \rangle $ as a fixed number measuring the contribution of $E$ to the momentum of the nucleon. The correct way to extract a measure of the separate contributions is to remember, as stressed by Jaffe and Manohar \cite{Jaffe:1989jz} in the QCD case,  that constituent quark models and parton  models of the nucleon are canonical Fock-space models. Thus
the physical nucleon states of the theory are taken to be superpositions of quark and gluon Fock states. Similarly, in QED, atomic states are regarded as superpositions of electron and photon Fock states. How this affects the extraction of the separate momentum and angular momentum contributions is spelled out in Section~\ref{Sep}. \nl
We shall argue that the various prescriptions given by Chen et.al. and Wakamatsu are somewhat \textit{ad hoc} and that what is missing is a compelling criterion for identifying a particular operator as the momentum operator or as the angular momentum operator. The natural definition  of the \emph{total} momentum operator is as the generator of translations and of the \emph{total} angular momentum operator as the generator of rotations, but when the system consists of different interacting quanta some modification is unavoidable. We suggest that \emph{the minimal requirement for this identification is the following}: \nl
\textbf{Definition}:  Suppose we have a system consisting of interacting fields $\phi_E(x)$ and $\phi_F(x)$. Then the \emph{momentum operator} $P^j_E(t)$ for, say, particles $E$
 should, \emph{at equal times}, satisfy

\beq \label{Eq:goodP} i[P^j_E(t)\, , \phi^E(t,\bm{x})]= \partial^j\phi^E(t, \bm{x}) .\eeq

Analogously, the \emph{ angular momentum operator} $M^{ij}_E(t) $ should, \emph{at equal times}, satisfy

\beq \label{Eq:goodM}  i[M^{ij}_E(t)  \,, \phi^E_r(t,\bm{x})] = (x^i\partial^j - x^j\partial^i)\phi^E_r(t,\bm{x}) +(\Sigma^{ij})_r^{\, \,s}\phi^E_s(t,\bm{x}) \eeq

where $r$ and $s$ are spinor or Lorentz labels and $(\Sigma^{ij})_r^{\, \,s}$ is the relevant spin operator. The need for the requirement ``\emph{at equal times}" is explained in detail in Section~\ref{Sep}. \nl
Demanding that these conditions  be satisfied leads to the conclusion that  the \emph{canonical} expressions for the   momentum and angular momentum operators  are  the correct and physically meaningful ones.
It is then an inescapable fact that the photon and gluon  angular momentum operators cannot, in general, be split in a gauge invariant way into a spin and orbital part. However, as discussed in Section~\ref{helicity}, the projection of the photon and gluon spin onto their direction of motion i.e. their helicity, is gauge invariant and is measured in deep inelastic scattering on atoms or nucleons respectively. \nl
It should  be noted that Ji's expressions for the components of the quark and gluon momentum and angular momentum vectors, which are the Bellinfante versions, do not conform to the above definition and thus should not be considered as measuring \emph{all} the components of the  physical quark and gluon momentum and angular momentum vectors, though it turns out that they give the correct results for the $Z$-components, $P_z$ and $J_z$, for a nucleon moving in the $Z$ direction. In particular the quark orbital angular momentum \emph{defined} by Ji as the difference between his quark total angular momentum, as measured in Deeply-virtual Compton Scattering, and the quark spin, as measured in Polarized DIS, is in agreement with our definition, as long as it is appreciated that this refers only to the components along the direction of motion of the nucleon.\nl

The difficulty in defining separate quark and gluon angular momenta in QCD has its analogue in QED, in the problem of defining separate electron and photon angular momenta. However, the situation is not completely analogous in the two cases, because the straightforward gauge invariance of QED is replaced by the rather different BRST \cite{Becchi:1975nq,*Tyutin} invariance of QCD. For this reason we shall discuss the two cases separately.

Most of the  problems which beset the definition of separate quark and gluon angular momenta actually already occur at the level of the linear momentum. Since this is a much simpler object to deal with, we shall mainly illustrate the problematic issues through an analysis of the linear momentum operator.

\section{\label{Observ} Observables in gauge theories}

As mentioned above we think there has been too much emphasis on the need to use gauge invariant operators to represent any dynamical quantity which can be measured i.e which is an \emph{observable}. In this section we shall show that, in fact, in gauge theories the concept of an observable is very subtle and is rather different in QED and QCD, and we shall give the precise conditions that an observable operator must satisfy. Our discussion follows the approach of Kugo and Ojima \cite{Kugo:1979gm}, which, in turn, follows the treatment of Strocchi and Wightman \cite{Strocchi:1974xh}.

In the covariant quantization of a gauge theory it is unavoidable that one has to set up the theory in a vector space with an indefinite metric i.e. one in which the ``length" or norm of a vector can be negative. From this one constructs a subspace, the physical vector space $\mathcal{V}_{phys}$, in which scalar products are positive semi-definite, and finally the positive definite Hilbert quotient space $\mathcal{H}_{phys}= \mathcal{V}_{phys}/\mathcal{V}_0$, where $\mathcal{V}_0$ is the subspace of $\mathcal{V}_{phys}$ consisting of zero-norm vectors\footnote{Strictly speaking $\mathcal{H}_{phys}$ should be defined as the \emph{completed} quotient space, but this is irrelevant for our discussion.}. How the states forming $\mathcal{V}_{phys}$  are defined, depends on the formulation of the theory. In covariantly quantized QED they are defined by $B^{(+)}(x)| \Phi \rangle = 0 $, where $B(x)$ is the gauge fixing field. In covariantly quantized QCD one has $Q_B| \Phi \rangle = 0 $ , where $Q_B$ is the generator of BRST transformations.    \nl

Note that while $\mathcal{V}_{phys}$ is labelled ``phys", the states which correspond to the actual physical particles belong to $ \mathcal{H}_{phys}$ i.e. the zero-norm states in $\mathcal{V}_{phys}$ are not truly physical. We assume, as usual, that the physical states form a complete set in $ \mathcal{H}_{phys}$.

 Let $| \Phi \rangle $ be a state in $\mathcal{V}_{phys}$ i.e. $| \Phi \rangle \in \mathcal{V}_{phys}$ and let $|\chi\rangle $ be a zero-norm state i.e. $|\chi\rangle \in \mathcal{V}_0 $. Then it can be shown that

\beq \label{Eq:Ortho} \langle \Phi | \chi \rangle =0  \qquad \textrm{for} \qquad \forall | \Phi \rangle \in \mathcal{V}_{phys}, \quad \forall | \chi \rangle \in \mathcal{V}_0 \eeq

i.e.

\beq \label{Eq:orthogspace}  \mathcal{V}_0 \perp \mathcal{V}_{phys}. \eeq

Let $O$ be a physical quantity and let $\hat{O}$ be the hermitian operator representing it. It can be shown that a necessary condition for $\hat{O}$ to be an observable is

\beq \label{Eq:defObs} \langle  \Phi + \chi   \,    |\, \hat{O} \, |   \Phi + \chi  \,    \rangle = \langle  \Phi \,    |\, \hat{O} \, |   \Phi \,    \rangle  \qquad \textrm{for} \qquad \forall | \Phi \rangle \in \mathcal{V}_{phys}, \quad \forall | \chi \rangle \in \mathcal{V}_0 \eeq

Equivalently, via Eq.~(\ref{Eq:orthogspace}), an observable operator must satisfy

\beq \label{Eq:obsOp}   \hat{O}\, |  \Phi  \rangle \in \mathcal{V}_{phys} \qquad \textrm{for} \qquad \forall |  \Phi  \rangle \in \mathcal{V}_{phys} \eeq

The essential point of this argument, as we shall see later, is that the condition Eq.~(\ref{Eq:obsOp}) does \emph{not} necessarily require an observable operator to be gauge invariant in the operator sense i.e. to commute with the generator of gauge transformations.  And we shall see that the situation differs somewhat between  covariantly quantized QED and QCD.

\section{\label{Totmom} The momentum operator in  gauge-invariant theories}

If the theory is invariant under translations in space-time, then Noether's theorem allows the construction, from the classical Lagrangian, of what is usually referred to as the canonical energy-momentum tensor density $t^{\, \mu  \nu}_{can}(x)$. This is a conserved density

\beq \label{Eq:t} \partial_\mu  t^{\, \mu \nu}_{can}(x) = 0 \eeq

but is generally not symmetric under $\mu \leftrightarrow \nu $.

The canonical \emph{total} linear momentum operator $P^j_{can}$ is the space integral

\beq \label{Eq:PC} P^j_{can}= \int d^3x \,\, t^{\, 0  j}_{can}(x) \eeq

and, crucially, is independent of time as a consequence of Eq.~(\ref{Eq:t}).

\subsection{The canonical momentum operator as generator of translations}

In the classical theory $P^j_{can}$ thus constructed is the generator of spatial translations. In the quantum theory one has to check that the operator version of
$P^j_{can} $ satisfies the correct commutation relations with all the fields i.e. for any field $\phi(x)$

\beq  \label{Eq:Pcomm} i\, [P^j_{can},\, \phi(x) ] = \partial^j \, \phi(x)  \eeq

It is important to realize that in an interacting field theory an arbitrary commutation relation between the fields cannot be calculated unless one can completely solve the theory---an impossible task in  all relevant physical theories. On the other hand the Equal Time Commutators (ETC) are fixed as part of the process of quantizing the theory. Hence the only reason it is possible to check an equation like (\ref{Eq:Pcomm}) is because $P^j_{can}$ is independent of time and so the time variable in the fields occurring in it can  be chosen to coincide with the time variable in $\phi(x)\equiv \phi(t, \bm{x})$. This consideration will play a crucial role when we come to discuss how to divide the total momentum into contributions from the different fields in the theory.

An important issue in comparing the treatment of linear and angular momentum is the concept of a \emph{local operator}. An operator $O(x)$  is local if, obviously, it is defined at one space-time point $x$, but also it must satisfy the law of translation

\beq \label{Eq:locOp} O(t,\bm{x}+\bm{a}) = e^{iP^j_{can} \, a_j} \, O(t,\bm{x}) \, e^{-iP^j_{can} \, a_j}.  \eeq

Note that an operator of the form $M(x)=xO(x)$, such as occurs in the expression for the angular momentum, is \emph{not} a local operator. (It is trivial to see that if $M(x)$ satisfies Eq.~(\ref{Eq:locOp}) then $M(x)=0$ for all $x$.) In a careful discussion of the properties of  angular momentum,  operators of this type have been called \emph{compound} operators \cite{Bakker:2004ib}.

\subsection{The Bellinfante energy momentum operator tensor density}

As mentioned the canonical $t^{\, \mu  \nu}_{can}(x)$ is generally not symmetric under interchange of $\mu$ and $\nu$. It is also not gauge invariant. It is possible to construct from $t^{\, \mu  \nu}_{can}(x)$ and the Lagrangian, the conserved Bellinfante density $t^{\, \mu  \nu}_{bel}(x)$, which is symmetric and, which is, in some cases, as will be discussed below,  gauge invariant. It differs from $t^{\, \mu  \nu}_{can}(x)$ by a divergence term of the following form:

\beq \label{Eq:Bell} t^{\, \mu \, \nu}_{bel}(x) = t^{\, \mu \, \nu}_{can}(x)  + \frac{1}{2}\partial_{\rho} [H^{\rho \mu \nu } -H^{\mu \rho \nu } - H^{ \nu \rho \nu }] \eeq

where the only relevant property of $H^{\rho \mu \nu }$ for the present discussion is that it is antisymmetric under $\mu \leftrightarrow \nu $

\beq \label{Eq:H}  H^{\rho \mu \nu } = - H^{\rho \nu \mu} \eeq
and that it is a local operator.

It follows that $P^j_{bel}$ defined by

\beq \label{Eq:PB} P^j_{bel} \equiv  \int d^3x \,\, t^{\, 0  j}_{bel}(x) \eeq

differs from $P^j_{can}$ by the integral of a spatial divergence, and it is usually stated that since the fields must vanish at infinity, such a contribution can be neglected, leading to the  equality

\beq \label{Eq:PBC}  P^j_{bel} = P^j_{can} . \eeq

Now for a classical \emph{c-number} field it is meaningful to argue that the field vanishes at infinity and that Eq.~(\ref{Eq:PBC}) holds as a numerical equality. It is much less obvious what this means for a quantum operator. The correct way to tell whether a divergence term can be neglected is to check what its role is in the relevant physical \emph{matrix elements} involving the operator. In the case of Eq.~(\ref{Eq:PBC}) one can readily check that the matrix elements between \emph{any} normalizable physical states, $| \Psi \rangle $ and $| \Phi \rangle $ are the same\footnote{This is not true for all operators which differ by a divergence term. Singularities can affect the result.} i.e.

\beq \label{Eq:ExpVal}   \langle  \Phi |\, P^j_{bel} \, |  \Psi \rangle = \langle  \Phi |\, P^j_{can}  \,|  \Psi \rangle . \eeq

However, the operators cannot be identical, because  one, for example, may be gauge invariant and the other not, so that the equality would be contradicted upon performing a gauge transformation. On the other hand the operators are essentially equivalent, and they generate the same  transformations on the fields. We shall indicate the relationship as

\beq \label{Eq:defEq} P^j_{bel} \cong   P^j_{can}. \eeq

It should be noted that it would be impossible to construct a consistent theory if it were not permissible, in certain case, to ignore the spatial integral of the divergence of a local operator. For example we could not even establish the obvious requirement that the momentum operator commutes with itself! For one has, (no sum over $j$)
\beq \label{Eq:consist} i[P^j\,,\,P^j] = \int d^3x \,\,i[P^j\,,\, t^{\, 0  j}(x)]= \int d^3x \,\, \partial^j t^{\, 0  j}(x)\eeq
and this vanishes only if the divergence integral can be ignored.

For compound operators like the  angular momentum it is a much more difficult task to show the equivalence of the \emph{total} angular momentum generators $M^{ij}_{can}$ and $M^{ij}_{bel}$, constructed from the  canonical and Bellinfante  $P_{can,bel}$ respectively, and care has to be exercised to always use normalizable states. This has been done by Shore and White \cite{Shore:1999be}.

\section{\label{QED} Quantum Electrodynamics}

We shall study the questions of gauge invariance and Lorentz covariance first in the simpler context of QED.

\subsection{The non-gauge invariance of  the QED momentum and angular momentum operators}

We remarked in the Introduction that in trying to define separate quark and gluon angular momentum operators too much emphasis was being placed on the use of gauge invariant operators by  Ji, Chen et al and Wakamatsu.

In support of this point of view we shall now prove  that in any theory which is invariant under a local c-number gauge transformation, even the   \emph{total} momentum and angular momentum operators \emph{cannot} be gauge invariant. As discussed in Section~\ref{Observ} this does \emph{not} mean that the momentum and angular momentum are not observables i.e. cannot be measured. Because what one measures are not operators but matrix elements of operators, and if care is exercised in defining the physical states of the theory (respecting any subsidiary conditions) then these matrix elements turn out to be gauge invariant.

\textbf{Theorem 1}: Consider a theory  which is invariant under local c-number gauge transformations. Let $P^\mu$ be the total  momentum operators, defined as the generators of space-time translations, and let $M^{ij}$ be the  total angular momentum operators, defined as the generators of rotations. Then  $P^\mu$ and $M^{ij}$ cannot be gauge invariant operators. \nl
\textbf{Proof}: For simplicity we consider QED and give the proof just for the momentum operators. The case of angular momentum is a   straightforward generalization. Note that it  is irrelevant for the proof whether we  use the canonical or Bellinfante versions. \nl
The theory is invariant under the infinitesmal gauge transformation

\beq  \label{ GT1  } A^\mu(x) \rightarrow A^\mu(x) +  \partial^\mu \Lambda (x) \eeq

where $\Lambda (x)$ is a c-number field satisfying $\Box \Lambda(x) = 0 $ and vanishing at infinity.

Now gauge transformations are canonical transformations \cite{Jauch:1955jr}. Let $F$ be the generator of gauge transformations, so that

\beq \label{GT3} i[F,A^\mu(x)] =  \partial^\mu \Lambda (x) \eeq

and consider the Jacobi identity

\beq \label{JI} [F,\, [P^\mu, A^\nu]] + [A^\nu,\, [F, P^\mu]] + [P^\mu,\, [A^\nu, F]] = 0 \eeq

Now $[P^\mu,\, [A^\nu, F]] = 0 $ since by Eq.~(\ref{GT3}) $[A^\nu, F]$ is a c-number and thus commutes with $P^\mu$, so that

\beq \label{JI2} [[F,P^\mu], \, A^\nu] = [F, \, [P^\mu, A^\nu]] \eeq

Moreover since $P^\mu$ are the generators of translations

\beq \label{Trans} i[P^\mu, A^\nu]= \partial^\mu A^\nu \eeq

Thus the RHS of Eq.~(\ref{JI2}) becomes

\beq \label{JI3} [F, \, [P^\mu, A^\nu]]=-i\partial^\mu [F, A^\nu(x)] = -\partial^\mu \partial^\nu \Lambda(x) \neq 0 \eeq

and hence from Eq.~(\ref{JI2})

\beq \label{JI4} [[F,P^\mu], \, A^\nu]\neq 0 \eeq

implying that

\beq \label{JI5} [F,P^\mu] \neq 0 \eeq
so that $P^\mu$ is not gauge invariant.

 \subsection{The momentum and angular momentum in QED are observables}

 We shall now demonstrate that this lack of gauge invariance is of no physical significance. We shall take as an example covariantly quantized QED and show that the matrix element of $P^j_{can}$ between any physical states, is unaffected by gauge changes in the operator.

As far as we are aware the most general covariantly quantized version of QED is given by the Lautrup-Nakanishi Lagrangian density \cite{Lautrup, Nakanishi:66}, which is a combination of the Classical Lagrangian ($Clas$) and a Gauge Fixing part ($Gf$)

 \beq \label{Eq:LN} {\cal{L}} = {\cal{L}}_{Clas}  +  {\cal{L}}_{Gf} \eeq

 where

 \beq \label{Eq:LClas} {\cal{L}}_{Clas} = -\frac{1}{4}F_{\mu\nu}F^{\mu\nu}
  +  \frac{1}{2}\, [\bar{\psi} (i\slas{\partial} - m + e \not\negthickspace{A} )\psi +  \textrm{h.c.}]  \eeq

  and

 \beq \label{Eq:Gf} {\cal{L}}_{Gf} = B(x)\,\partial_{\mu}A^{\mu}(x)+ \frac{\textsf{a}}{2}B^2(x) \eeq

  where $B(x)$ is the gauge-fixing field\footnote{Because of its similarity with the QCD case, we use the notation of Nakanishi. Note that Lautrup's $\Lambda(x)=-B(x)$.} and the parameter $ \textsf{a} $ determines the structure of the photon propagator and is irrelevant for the present discussion\footnote{The case $\textsf{a}=1$ corresponds to the Gupta-Bleuler approach (see e.g. \cite{Jauch:1955jr}) based on the Fermi Lagrangian.  }. The theory is invariant under the usual c-number infinitesmal gauge transformation

  \beq \label{Eq:GT} A_\mu \rightarrow A_\mu + \partial_{\mu}\Lambda(x) \qquad \psi \rightarrow \psi +ie\Lambda \psi \eeq
 while $B(x)$ is taken to be unaffected by gauge transformations.

  A straightforward calculation gives for the  conserved generator of infinitesmal gauge transformations

  \beq  \label{Eq:F} F = -\,\int d^3 x \, [e\bar{\psi}\gamma^0\psi \Lambda(x) + F^{0j}\partial_j\Lambda(x) - B(x)\partial_0\Lambda(x) ] \eeq

  which,via the equations of motion, can be transformed to

  \beq \label{Eq:Ftr} F = \int d^3 x \, [(\partial_0 B)\Lambda - B\partial_0\Lambda +\partial_j(F^{0j}\Lambda)]. \eeq

  Now the \emph{physical} states $|\Phi \rangle $ of the theory are defined to satisfy

  \beq \label{Eq:phys} B^{(+)}(x) | \Phi \rangle = 0  \eeq

  where

  \beq \label{Eq:B} B(x)  =B^{(+)}(x) +   B^{(-)}(x)  \eeq

  with $B^{(\pm)}(x)$  the positive/negative frequency parts of $B(x)$.

  With this definition of the physical states, an operator $\hat{O}$ is an observable, if, according to Eq.~(\ref{Eq:defObs}), $\hat{O}|\, \Phi \,\rangle$ is itself a physical state i.e. if
  \beq \label{Eq:defObsQED} B^{(+)}(x)\, (\hat{O}|\, \Phi \,\rangle ) = 0. \eeq

  This is equivalent to the condition
  \beq \label{Eq:defObsQED'}      [\,B^{(+)}(x),\,\hat{O}]\,|\, \Phi \,\rangle  = 0 \eeq
  since
  \beqy \label{Eq:comQED} [\,B^{(+)}(x),\,\hat{O}]\,|\, \Phi \,\rangle &= & B^{(+)}(x)\,\hat{O}|\, \Phi \,\rangle - \hat{O}\,B^{(+)}(x)|\, \Phi \,\rangle =B^{(+)}(x)\,\hat{O}|\, \Phi \,\rangle \nn \\
  &=& 0 \qquad \textrm{iff Eq.~(\ref{Eq:defObsQED}) holds.} \eeqy
 Since, via  Eqs.~(\ref{Eq:phys}) and (\ref{Eq:Pcomm})
 \beq \label{Eq:PQEDObs}  [B^{(+)}(x), P^j]\, |\Phi \rangle = i\partial^jB^{(+)}(x)\, |\Phi \rangle = 0  \eeq
 we see that $P^j$ is an observable, so that its eigenstates are physical states.

 We shall now consider the gauge invariance of its matrix elements. In doing so it should be noted
 that $B^{(-)}(x)=[B^{(+)}]^{\dag}(x)$, so that
  $ \langle \Phi | B^{(-)}(x) =0$,  and thus  for arbitrary physical states
\beq \label{Eq:MelB}  \langle \Phi' | B(x) |\Phi \rangle  =0 . \eeq

  \textbf{Theorem 2} Any physical matrix element of the momentum operator $P^j$ is invariant under gauge transformations. \nl
  \textbf{Proof} Consider the general physical matrix element

  \beq \langle \Phi' | P^j |\Phi \rangle = \int d^3 \bm{p} \, d^3 \bm{p}' \, \phi'^*(\bm{p'}) \, \phi(\bm{p})\, \langle \bm{p}' |P^j | \bm{p} \rangle \eeq

  The change induced in $\langle \bm{p}' |P^j | \bm{p} \rangle $ by the gauge transformation is given by $\langle \bm{p}' |i [ F,P^j] | \bm{p} \rangle $. Focus initially on the effect of the first two terms (call them $f_{12}$)  in the integrand on the RHS of  Eq.~(\ref{Eq:Ftr}).

  \beq \label{Eq:f12} \langle \bm{p}' |i [ f_{12},P^j] | \bm{p} \rangle = (p - p' )^j\, \langle \bm{p}' |f_{12} | \bm{p} \rangle = 0 \eeq
  because of Eq.~(\ref{Eq:MelB}) and the fact that $\Lambda$ is a c-number.

  The change induced by the third, divergence term (call it $f_3$)  in the integrand  on the RHS of  Eq.~(\ref{Eq:Ftr}), after some algebra, and using translation invariance Eq.~(\ref{Eq:locOp}), can be written

  \beqy \label{Eq: f3} \langle \bm{p}' |i [ f_3,P^j] | \bm{p} \rangle &= &(p'-p)^j \{(p -p')^0\, \langle \bm{p}' |A^k(0) | \bm{p} \rangle  \\
  &- & (p -p')^k \, \langle \bm{p}' |A^0(0) | \bm{p} \rangle\} \, \partial_k [\Lambda(x)\,e^{i(p-p')\centerdot x}] \eeqy

  and this vanishes after the spatial integration because $\Lambda(x)$ vanishes at infinity.

  Hence $\langle \Phi' | P^j |\Phi \rangle$ is indeed invariant under gauge transformations.

  \textbf{Corollary} Physical matrix elements of the total angular momentum operator $J^k$ are gauge invariant.

 The total angular momentum is given by
  \beq \label{Eq:defJ}  J^k = \frac{1}{2}\epsilon_{klm}\,M^{lm} = \frac{1}{2}\epsilon_{klm}\,\int d^3x {\cal{M}}^{0lm}(x)  \eeq

  where ${\cal{M}}^{0lm}(x)$ is the angular momentum tensor density. The simplest way to show the gauge invariance of the physical matrix elements in this case is to reinterpret the gauge change $i[F,M^{lm}]$ as $-i[M^{lm},F]$ and to study the effect of the rotations on $F$. For this one needs the following results:
  \beq \label{Eq:rotA} i[M^{lm},A^\beta (x)] = (x^l\partial^m -x^m\partial^l)A^\beta (x) + g^{l\beta} A^m(x) - g^{m\beta}A^l(x) \eeq
  and, since $B(x)$ is a scalar field,
  \beq \label{Eq:rotB} i[M^{lm},B(x)] = (x^l\partial^m -x^m\partial^l)B(x) \eeq
  Application of these to $F$ yields terms which either vanish directly as a result of the subsidiary condition Eq.~(\ref{Eq:MelB}) or divergence terms which can be shown to vanish since $\Lambda(x)$ vanishes at infinity.

The fact that even the \emph{total} momentum and angular momentum are not gauge invariant, but that their physical matrix elements are,  suggests that to insist on gauge-invariant operators for the momentum and angular momentum operators of the individual fields of the theory is unnecessary.

\subsection{Relativistic covariance in QED}

In the debate with Chen et al, Ji rightly argues that their photon vector potential does not transform as a 4-vector under Lorentz transformations, and implies that in his treatment his $A_\mu(x)$ transforms as a true 4-vector, and that this is an essential property. But if this is the case then Ji's expressions for momentum and angular momentum are incomplete. The point is that the gauge-fixing field $B(x)$ introduced above in the covariant quantization of QED also appears in the expressions for the momentum and angular momentum. One finds for the conserved canonical energy momentum tensor density,
\beq \label{Eq:canQED} t^{\mu\nu}_{can}=   \theta^{\mu\nu}_{can}     + t^{\mu\nu}_{can}(Gf) \eeq
where

\beq \label{Eq:thetacQED} \theta^{\mu\nu}_{can} = \frac{i}{2}\, \bar{\psi}\gamma^\mu \overleftrightarrow{\partial}^\nu \,\psi - F^{\mu\beta}\partial^\nu A_\beta  - g^{\mu\nu}{\cal{L}}_{Clas} \eeq

where $ \overleftrightarrow{\partial}^\nu \equiv \overrightarrow{\partial}^\nu - \overleftarrow{\partial}^\nu $, and

\beq \label{Eq:tcanGf} t^{\mu\nu}_{can}(Gf) = B \partial^\nu A^\mu - g^{\mu\nu}{\cal{L}}_{Gf}. \eeq
For the conserved Bellinfante density one finds,

\beq \label{Eq:tbQED}  t^{\mu\nu}_{bel}  = \theta^{\mu\nu}_{bel} + t^{\mu\nu}_{bel}(Gf) \eeq

where $\theta^{\mu\nu}_{bel}$, which is referred to as the classical energy momentum tensor density, is

\beq \label{Eq:tClas} \theta^{\mu\nu}_{bel} =  \frac{i}{4}\, \bar{\psi}(\gamma^\mu \overleftrightarrow{D}^\nu + \gamma^\nu \overleftrightarrow{D}^\mu )\,\psi - F^{\mu\beta}F^\nu_{\phantom{\nu}\beta} - g^{\mu\nu}{\cal{L}}_{Clas} \eeq

where $\overleftrightarrow{D}^\nu = \overleftrightarrow{\partial}^\nu -2ieA^\nu $,
and

\beq \label{Eq:tGf}  t^{\mu\nu}_{bel}(Gf) = - (\partial^\mu B)\,A^\nu -(\partial^\nu B)\,A^\mu - g^{\mu\nu}{\cal{L}}_{Gf}  \eeq

The conservation of an energy momentum tensor depends on the equations of motion, which are a consequence of the Lagrangian. Thus $
t^{\mu\nu}_{bel}$ is conserved, but $\theta^{\mu\nu}_{bel}$ is not, when the Lagrangian is ${\cal{L}}_{Clas}  +  {\cal{L}}_{Gf}$. On the other hand $\theta^{\mu\nu}_{bel}$ would be conserved \emph{if} the Lagrangian were ${\cal{L}}_{Clas}$.

Now Ji and Chen at al utilize $\theta^{\mu\nu}_{bel}$ and treat it as if it were conserved i.e. they take the momentum operator based on it
to be independent of time (equivalently: to remain unrenormalized), which implies that the Lagrangian is just ${\cal{L}}_{Clas}$. But it is well known that one cannot quantize QED \emph{covariantly} using ${\cal{L}}={\cal{L}}_{Clas}$. Nonetheless Ji insists that his $A_\mu$ transforms covariantly, which is thus, at the operator level, a contradiction.

We have seen that insisting on covariant quantization leads to a more  complicated structure for the energy momentum density and analogously for the angular momentum density. This raises what, at first sight, seems to be a worrying issue concerning several papers in the literature, e.g.
   Ji \cite{Ji:1996ek,Ji:1996nm,Ji:1997pf},  Jaffe and Manohar \cite{Jaffe:1989jz},  Bakker, Leader and Trueman (BLT) \cite{Bakker:2004ib} and Wakamatsu \cite{Wakamatsu:2010qj,Wakamatsu:2010cb}, where the general structure of the matrix elements of $\theta^{\mu\nu}_{bel}$ (or its QCD analogue) is derived under the assumption that $\theta^{\mu\nu}_{bel}$ is a genuine conserved tensor. However the situation is saved by the following: for  physical matrix elements, for both the canonical and Bellinfante versions,

   \beq \label{Eq:NonCont} \langle  \Phi' |\,t^{\mu\nu}(Gf)\,| \Phi \rangle =0. \eeq

   This follows from Eqs.~(\ref{Eq:tcanGf}, \ref{Eq:tGf}) and (\ref{Eq:Gf}) when a complete set of physical states is inserted between the operators appearing in $t^{\mu\nu}(Gf)$ and use is made of Eq.~(\ref{Eq:MelB}). Hence

   \beq \langle  \Phi' |\,\partial_{\mu} \theta^{\mu\nu}_{bel}(x)\,| \Phi \rangle = \langle  \Phi' |\,\partial_{\mu} t^{\mu\nu}_{bel}(x)\,| \Phi \rangle =0. \eeq

   Similar arguments show that $\theta^{\mu\nu}_{can}(x)$ , which just corresponds to the canonical version of $\theta^{\mu\nu}_{bel}(x)$,
may also be treated as a conserved density inside physical matrix elements.
  Thus this aspect of the analysis in the above papers is, in fact, consistent.

   In summary covariant quantization of QED complicates some aspects and there is no compelling reason to insist on it. Indeed, as explained in the Introduction, the non-covariant Coulomb gauge leads to a perfectly good Lorentz invariant theory. However, if one prefers to work with a covariantly quantized theory then, in so far as its physical matrix elements are concerned, $\theta^{\mu\nu}_{bel}(x)$  and $\theta^{\mu\nu}_{can}(x)$ may be treated as  conserved tensor operators.

 \section{\label{QCD} Quantum Chromodynamics}

 The situation in QCD is somewhat different.The infinitesmal gauge transformations on the gluon vector potential and on the quark fields, under which the pure quark-gluon Lagrangian ${\cal{L}}_{qG}$ ( the QCD analogue of the QED ${\cal{L}}_{Clas})$,

 \beq \label{Eq:LqG} {\cal{L}}_{qG}= -\frac{1}{4}G^a_{\mu\nu}G^{\mu\nu}_a + \frac{1}{2}\bar{\psi}^l[\delta_{lm}\,i\,(\overrightarrow{\slas{\partial}} -
\overleftarrow{\slas{\partial}} ) - 2 \,g t^a_{lm}\, \not{\negthickspace}{A}^a] \psi^m \eeq

is invariant, are determined by eight scalar c-number fields $\theta^a(x)$,

  \beq \label{Eq:GluonGT} \delta \,A^a_\mu = \partial_\mu \theta^a(x) - gf_{abc}A_b^\mu(x) \theta^c(x) \eeq

  \beq \label{Eq:quarkGT} \delta \psi^l=-igt^a_{lm}\theta^a(x)\psi^m(x) \eeq

 where $a,b,c =1,2...8$  and $l,m=1,2,3$ are colour labels, and where our sign convention is

 \beq \label{Eq:GQCD} G^a_{\mu\nu} =\partial_\mu A^a_{\nu} - \partial_\nu A^a_{\mu} -gf{abc}A^b_{\mu}A^c_{\nu}. \eeq

  However, in order to quantize the theory covariantly one has to introduce both a gauge-fixing field $B(x)$ and Fadeev-Popov anti-commuting fermionic ghost fields $c(x), \,\bar{c}(x)$.  The Kugo-Ojima Lagrangian \cite{Kugo:78}  for the covariantly quantized theory is then

  \beq \label{Eq:KO} {\cal{L}} = {\cal{L}}_{qG} + {\cal{L}}_{Gf+Gh} \eeq

  where

  \beq \label{Eq:Gf+Gh} {\cal{L}}_{Gf+Gh} =  -
  i(\partial^\mu\bar{c}^a)D^{ab}_\mu c_b  - (\partial^\mu B^a) A^a_\mu + \frac{\textsf{a}}{2}B^aB^a  \eeq

 which is no longer invariant under the original infinitesmal gauge transformations Eqs.~(\ref{Eq:GluonGT}, \ref{Eq:quarkGT}).

One can again show that the momentum operators $P_{can}, P_{bel}$ are not gauge invariant, but this is now irrelevant, given that the Lagrangian itself does not possess this invariance. Instead the theory is invariant under the BRST transformations \cite{Becchi:1975nq,*Tyutin}

\beqy \label{Eq:BRST} \delta A^a_\mu &=& \theta D^{ab}_\mu c^b(x) \nn \\
\delta \psi^l &=& -i\theta g  t^a_{lm}c^a(x)\psi^m(x) \nn \\
\delta c^a(x) &=& \theta (g/2) f_{abc}c^b(x)c^c(x) \nn \\
\delta \bar{c}^a  &=&  i\theta B^a(x) \nn \\
\delta B(x)& =& 0. \eeqy

where $\theta$ is a constant operator which commutes with bosonic fields and anti-commutes with fermionic fields.

The BRST transformation is generated by $\theta Q_B$ i.e.  for any of the above fields $\phi$

\beq \label{Eq:QBtr} i[\theta Q_B, \phi] = \delta \phi \eeq

where the conserved, hermitian charge $Q_B$ is given by

\beq \label{Eq:QBdef} Q_B= \int d^3x [ B^a {\overleftrightarrow{\partial}}_0c^a -gB^af_{abc}A^b_0c^c -i(g/2) (\partial_0\bar{c}^a) f_{abc}c^bc^c]. \eeq

There is also a conserved charge

\beq \label{Eq:Qc} Q_c=\int d^3x [\bar{c}^a\overleftrightarrow{\partial}_0 c^a -g \bar{c}^af_{abc}A^b_0c^c] \eeq

which ``measures" the \emph{ghost number}

\beq \label{Eq:GNo} i[Q_c, \phi] = N \phi \eeq

where $N=1$  for  $\phi=c,\,\, -1 \,\, \textrm{for}\,\, \phi=\bar{c} \,\, \textrm{and} \,\, 0$   for all other fields.

The physical states $|\Psi \rangle $ are defined by the subsidiary conditions
\beq \label{Eq:Phys} Q_B |\Psi \rangle =  0  \eeq
\beq  \label{Eq:Phys'} Q_c |\Psi \rangle  = 0 \eeq

\subsection{The momentum and angular momentum operators in covariant QCD}

The proof of an analogue of \textbf{Theorem 1} for BRST transformations does not work, because the BRST $\delta A^a_\mu$ is an operator, not a c-number. Consequently, use of the Jacobi identity Eq.~(\ref{JI}), with $F$ replaced by $Q_B$, does not imply that $P_{can}$ or $P_{bel}$ are non-BRST invariant.

Analogously to condition Eq.~(\ref{Eq:defObsQED'}), in order to be observable the momentum operator in QCD must satisfy

\beq \label{Eq:defObsQCD} [Q_B \, , P^j]|\Psi \rangle=0  \quad \textrm{and} \quad [Q_c \, , P^j]|\Psi \rangle=0. \eeq
The latter, as will be seen presently, follows from the fact that the ghost number of $P^j$ is zero. The former is usually stated to hold because $Q_B$ is a translationally invariant scalar. This is correct, but is not quite as trivial as it seems, for if we write
\beq \label{Eq:transInv} Q_B= \int d^3x \mathcal{Q}_B(t,\bm{x}) \eeq
then translational invariance requires
\beq \label{Eq:transInv'} e^{iP^ja_j} Q_B e^{-iP^ja_j} = \int d^3x e^{iP^ja_j} \mathcal{Q}_B(t, \bm{x}) e^{-iP^ja_j} =\int d^3x \mathcal{Q}_B(t, \bm{x} +\bm{a}) = Q_B. \eeq
The last step holds only if the integral in invariant under the change of variables $\bm{x}\rightarrow \bm{y}=\bm{x}+\bm{a}$, which is in accord with our being able to ignore the integral of a divergence.

One finds for the canonical energy momentum tensor density,
\beq \label{Eq:tcQCD} t^{\mu\nu}_{can}= t^{\mu\nu}_{can}(qG) + t^{\mu\nu}_{can}(Gf+Gh) \eeq

where
\beq \label{Eq:tcanqG} t^{\mu\nu}_{can}(qG) = \frac{i}{2}\bar{\psi}_l \gamma^\mu  \overleftrightarrow{\partial}^\nu  \psi_l - G^{\mu\beta}_a {\partial^\nu} A^a_\beta - g^{\mu\nu}{\cal{L}}_{qG} \eeq

and where
\beq \label{Eq:tcG} t^{\mu\nu}_{can}(Gf+Gh)= -A^\mu_a {\partial^\nu} B_a  -i({\partial^\nu} \bar{c}_a)(D^\mu_{ab}c_b)  - g^{\mu\nu} {\cal{L}}_{Gf+Gh} -i ({\partial^\mu}\bar{c}_a)({\partial^\nu}c_a). \eeq

The Bellinfante version is
\beq \label{Eq:tbQCD} t^{\mu\nu}_{bel}=  t^{\mu\nu}_{bel}(qG) + t^{\mu\nu}_{bel}(Gf+Gh) \eeq

where
\beq \label{Eq:tbelqG} t^{\mu\nu}_{bel}(qG)= \frac{i}{4}[\bar{\psi}_l \gamma^\mu \overleftrightarrow{D}^\nu \psi_l + (\mu \leftrightarrow \nu )] - G^{\mu\beta}_a G^\nu_{a\beta}- g^{\mu\nu}{\cal{L}}_{qG}  \eeq

is BRST invariant, i.e. commutes with $Q_B$. Here $\overleftrightarrow{D}^\nu $ is a matrix in colour space
\beq \label{DQCD} \overleftrightarrow{D}^\nu(z) = \delta_{lm}[\overrightarrow{\partial}^\nu - \overleftarrow{\partial}^\nu ] +2ig A_a^\nu(z)t^a_{lm}. \eeq
The gauge-fixing and ghost terms are given by
\beq \label{Eq:tbelGf} t^{\mu\nu}_{bel}(Gf+Gh)= - (A^\mu_a \partial^\nu B_a + A^\nu_a \partial^\mu B_a) -i[(\partial^\mu\bar{c}_a)D^{\nu}_{ab} c_b +(\partial^\nu\bar{c}_a)D^{\mu}_{ab} c_b ] - g^{\mu\nu} {\cal{L}}_{Gf+Gh}. \eeq

This can be rewritten \cite{Kugo:1979gm} as an anti-commutator with $Q_B$
\beq \label{tbelGfnew} t^{\mu\nu}_{bel}(Gf+Gh)= - \{ Q_B, \, \big( (\partial^{\mu} \bar{c}_a) A^{\nu}_a + (\partial^{\nu} \bar{c}_a) A^{\mu}_a + g^{\mu\nu}[\frac{ \texttt{a}}{ 2}\bar{c}_aB_a - (\partial^{\rho}\bar{c}_a) A_\rho^a ] \big) \}. \eeq

It follows that $t^{\mu\nu}_{bel}(Gf+Gh)$ is BRST invariant (because $Q_B$ is nilpotent i.e. $Q_B^2=0$) and does not contribute to physical matrix elements i.e.
\beq \label{tbelQCD} \langle  \Phi'   |\,  t^{\mu\nu}_{bel} \, |  \Phi \rangle = \langle  \Phi'   |\,  t^{\mu\nu}_{bel}(qG) \, |  \Phi \rangle. \eeq

Thus the entire $t^{\mu\nu}_{bel}(x)$ commutes with $Q_B$ and is therefore a \emph{local} observable.

The situation with $t^{\mu\nu}_{can}(x)$ is somewhat different. It does not commute with $Q_B$, so is not itself an observable, but, contrary to
 the statement in \cite{Shore:1999be},  $t^{\mu\nu}_{can}(Gf+Gh)$ does \emph{not} contribute to physical matrix elements. This can be seen as follows. The first three terms in Eq.~(\ref{Eq:tcG}) can be written as an anti-commutator with $Q_B$, so, as argued above, do not contribute to physical matrix elements. For the last term we have, by completeness,

 \beq \label{Eq:cc} -i \langle  \Phi'   |\,  ({\partial^\mu}\bar{c}_a)({\partial^\nu}c_a) \, |  \Phi \rangle = -i \sum_{all \, \Psi}\langle  \Phi'   |\,  ({\partial^\mu}\bar{c}_a) \, |  \Psi \rangle \langle  \Psi   |\,  ({\partial^\nu}c_a) \, |  \Phi \rangle .\eeq

 This is zero because, via Eq.~(\ref{Eq:GNo}), $ c_a(x)= i \, [Q_c, c_a(x)] $, so that

 \beq \label{Eq;cc'} \langle  \Psi   |\,  \partial^\nu c_a(x) \, |  \Phi \rangle = i \, \partial^\nu \langle  \Psi   |\, [Q_c, c_a(x)]  \, |  \Phi \rangle = 0 \eeq

 as a consequence of Eq.~(\ref{Eq:Phys'}).

 Thus even though the actual canonical \emph{density} is not BRST invariant, its ghost and gauge-fixing terms do not contribute to physical matrix elements. And,
as discussed in Section \ref{Totmom}~B, for the space integrated versions, because they differ by a divergence, we  have, analogous to Eq.~(\ref{Eq:PB}),

\beq \label{Eq:EqQCD}  P^j_{bel}(QCD)\cong P^j_{can}(QCD) \eeq
and both are BRST invariant.

There is thus no compelling reason in QCD for insisting on using the Bellinfante version.
 Analogous statements hold for the angular momentum generators $M^{ij}_{can}$ and $M^{ij}_{bel}$.

\subsection{Relativistic covariance in QCD}

We have seen that insisting on covariant quantization forces us to include gauge-fixing and ghost fields in the Lagrangian. However, the terms in the canonical and Bellinfante versions of the total momentum, which depend on the ghost and gauge-fixing fields, do not contribute to physical matrix elements. Thus if we consider the expectation value of the \emph{total} momentum operator for  a nucleon in a state of definite momentum $|\, \bold{p}\, \rangle$ then, irrespective of whether we use $P_{can}$ or $P_{bel}$, there will be no contribution from the ghosts or gauge-fixing fields. Moreover, both operators are observables and their matrix elements are thus physically measurable quantities.

 However, as mentioned in the Introduction,  Ji, Chen et al and Wakamatsu insist on using the gauge-invariant Bellinfante tensor, or modifications of it, for the \emph{separate} electron and photon, or quark and gluon, parts of the total momentum and angular momentum tensors.
  We shall argue in the next Section that this has no solid basis, is essentially arbitrary and lacks any persuasive physical motivation.

\section{\label{Sep} The problem of defining separate quark and gluon momenta}

We come now to the heart of the controversy between Ji,  Chen at al and Wakamatsu, namely how to define in a sensible way the separate contributions of quarks and gluons to the momentum and angular momentum of a nucleon. There are actually two separate issues. One, quite general, is how to define the separate momenta for a system of interacting particles. The second is more specific to gauge theories and includes the issue of splitting the angular momentum of a gauge particle into a spin and orbital part.

\subsection{Interacting particles: the general problem}

Suppose we have a system of interacting particles $E$ and $F$ and we split the total momentum into two pieces

\beq \label{Eq:split}  P^j=P^j_E +P^j_F  \eeq

which we wish to associate with the momentum carried by the individual particles $E$ and $F$ respectively.

As mentioned in the Introduction it is crucial to realize that Eq.~(\ref{Eq:split}), as it stands, is totally misleading, and should be written

\beq \label{Eq:Split} P^j=P^j_E(t) +P^j_F(t)  \eeq

to reflect the fact that the particles exchange momentum as a result of their interaction. \nl

The key question is: what should be the criterion for identifying $P_{E,F}$ as the momentum associated with particles $E,F$ respectively?

The seductively obvious answer would be to demand that

\beq \label{Eq:bad} i[P^j_E , \phi^E(x)]= \partial^j\phi^E(x) \eeq

and similarly for $F$, but there is no way we can check this, since $P^j_E(t)$ depends on $t$ and, without solving the entire theory, we are only able to compute equal time commutators .

 We suggest, therefore, that \emph{the minimal requirement for identifying an operator $P^j_E$ with the momentum carried by $E$} , is to demand that \emph{at equal times} the analogue of Eq.~(\ref{Eq:bad}) holds i.e.

\beq \label{Eq:goodP} i[P^j_E(t)\, , \phi^E(t,\bm{x})]= \partial^j\phi^E(t, \bm{x}) .\eeq

Analogously, \emph{for an angular momentum operator $M^{ij}_E$ we suggest the minimal requirement} that

\beq \label{Eq:goodM}  i[M^{ij}_E(t)  \,, \phi^E_r(t,\bm{x})] = (x^i\partial^j - x^j\partial^i)\phi^E_r(t,\bm{x}) +(\Sigma^{ij})_r^{\, \,s}\phi^E_s(t,\bm{x}) \eeq

where $r$ and $s$ are spinor or Lorentz labels and $(\Sigma^{ij})_r^{\, \,s}$ is the relevant spin operator.

Now we explained in Section~\ref{Totmom}B  that for the total momentum there is no essential difference between $P_{can}$ and $P_{bel}$, since their integrands differ by the spatial divergence of a local operator. However, if we split $P_{can}$ into $P_{can,\,E} + P_{can,\,F}$ and $P_{bel}$ into $P_{bel,\,E} + P_{bel,\,F}$, then the integrands of $P_{can,\,E}$ and $P_{bel,\,E}$  do \emph{not} differ by a spatial divergence, and hence  $P_{can,\,E}$ and $P_{bel,\,E}$  do not generate the same transformation on $\phi^E(x)$, and similarly for $F$.

As an example consider QED. From Eqs.~(\ref{Eq:canQED}, \ref{Eq:thetacQED}) and Eqs.~(\ref{Eq:tbQED}, \ref{Eq:tClas}, \ref{Eq:tGf}) we would identify

\beq  \label{Eq:canEl}   t^{0j}_{can}(\textrm{electron}) = \frac{i}{2}\, \bar{\psi}\gamma^0 \overleftrightarrow{\partial}^j \,\psi \eeq

and

\beq \label{Eq:belEl} t^{0j}_{bel}(\textrm{electron}) = \frac{i}{4}\, \bar{\psi}(\gamma^0 \overleftrightarrow{D}^j + \gamma^j \overleftrightarrow{D}^0 )\,\psi \eeq

and these do not differ by a spatial divergence.

It should be noted that the difference between various definitions of the momentum operators is not just a question of principle.
In QCD the asymptotic $(Q^2\rightarrow \infty )$ limit of the longitudinal momentum carried by quarks in a nucleon, with  the Ji definition is $P(\textrm{quarks})_{Ji} \approx 50\% $ whereas with the  Chen et al version  $P(\textrm{quarks})_{Chen}\approx 80\%$,  for the number of flavours $n_f=5$.

Since, by construction, $P_{can,\,E}$ and $P_{can,\,F}$ generate the correct transformations on $\phi_E(x)$ and $\phi_F(x)$ respectively, we conclude that with the above minimal requirement we are forced to associate the  momentum and angular momentum of $E$ and $F$ with the \emph{canonical version} of the relevant  operators. This  disagrees with  Ji, Chen et al and Wakamatsu, but agrees with Jaffe and Manohar \cite{Jaffe:1989jz}. \nl
 Nonetheless, exceptionally,  for the fraction of the $Z$-component of the longitudinal momentum and angular momentum carried by the quarks in a nucleon moving in the $Z$ direction, the distinction between Bellinfante and canonical versions is not crucial, since it turns out that $P_z(\textrm{quarks})_{Ji}\equiv P_z(\textrm{quarks})_{bel} = P_z(\textrm{quarks})_{can}$ and  $J_z(\textrm{quarks})_{Ji}\equiv J_z(\textrm{quarks})_{bel} =J_z(\textrm{quarks})_{can}$ , as will be discussed in Section~\ref{longMom}.

Now, as Jaffe and Manohar \cite{Jaffe:1989jz} have emphasized in the QCD case,  constituent quark models and parton models of the nucleon are canonical Fock-space models. Thus the physical states of the theory are taken to be superpositions of Fock states, formed from the vacuum by the quark and gluon ``in-field" creation operators. Similarly, in QED, atomic states are regarded as superpositions of Fock states, formed from the vacuum by the electron and photon ``in-field" creation operators.
 Loosely speaking, for any field $\phi(x)$ \footnote{Strictly speaking such limits of operators should be carried out using normalizable ``smearing functions". We shall continue to be a little cavalier in order not to complicate the presentation.}

\beq \label{Eq:inOp} \phi(x) \xrightarrow{t\rightarrow -\infty}   \sqrt{Z}\phi_{in}(x) \eeq
where $Z$ is a renormalization constant.
Also
\beqy \label{Eq:Pin} [P^j_{can,\,E}(t),\phi_E(t,\bm{x})]& & \xrightarrow{t\rightarrow -\infty}  [P^j_{in,\,can}(E) \, ,\sqrt{Z}\phi_{in,\,E}(t,\bm{x})]  \\
\partial^j \phi_{E}(t,\bm{x})       & &  \xrightarrow{t\rightarrow -\infty} \sqrt{Z} \partial^j \phi_{in,\,E}(t,\bm{x}) \eeqy
where we have defined
\beq \label{Eq:defPin}   P^j_{can,\,E}(t) \xrightarrow{t\rightarrow -\infty} P^J_{in,\,can}(E).\eeq
Note that because the ``in"  fields obey free field equations, $P^J_{in,\,can}(E)$ is independent of time.

Now as we have stressed $P_{E,F}(t)$ are time-dependent operators. However, these operators possess a remarkable property. While their general matrix elements are time-dependent, there is a sub-class of these, and it is just this class of matrix elements that are of interest to us, which are time-independent, namely, their matrix elements between arbitrary states of a \emph{single particle}. To see this for the momentum $P_E^j(t)$ let

\beq |\,\psi \,\rangle = \int d^3 p' \, \psi (\bm{p}') \, |\, \bm{p}' \,\rangle \quad \textrm{and} \quad |\,\phi \,\rangle = \int d^3 p \, \phi (\bm{p}) \, |\, \bm{p} \,\rangle \eeq
be arbitrary states of of a particle of mass m, so that
\beq p_0^2= \bm{p}^2 + m^2 \quad \textrm{and} \quad p_0'^2= \bm{p}'^2 + m^2. \eeq

Then
\beqy \langle \, \psi   \, |\, P_E^j(t) \, | \,   \phi    \, \rangle &= &\int d^3\bm{p}'\,d^3\bm{p} \,d^3\bm{x}\,\psi^*(\bm{p}')\,\phi(\bm{p}) \langle \, \bm{p}'   \, |\, t_E^{0j}(x) \, | \,   \bm{p}    \, \rangle  \nn \\
&=& \int d^3\bm{p}'\,d^3\bm{p} \,d^3\bm{x}\,\psi^*(\bm{p}')\,\phi(\bm{p})e^{i\bm{x}\centerdot(\bm{p}'-\bm{p})} \, e^{it(p_0-p'_0)} \langle \, \bm{p}'   \, |\, t_E^{0j}(0) \, | \,   \bm{p}    \, \rangle \nn \\
&=& (2\pi)^3 \,\int d^3\bm{p}'\,d^3\bm{p} \,\psi^*(\bm{p}')\,\phi(\bm{p}) \,\delta^3(\bm{p}' - \bm{p})\, e^{it(p_0-p'_0)} \langle \, \bm{p}'   \, |\, t_E^{0j}(0) \, | \,   \bm{p}    \, \rangle \nn \\
&=& (2\pi)^3 \,\int \,d^3\bm{p} \,\psi^*(\bm{p})\,\phi(\bm{p}) \, \langle \, \bm{p}   \, |\, t_E^{0j}(0) \, | \,   \bm{p}    \, \rangle \eeqy
which is independent of time because $p'_0=p_0=\sqrt{\bm{p}^2 + m^2} $. \nl

A similar, though more complicated argument, shows that the single particle matrix elements of the angular momentum operators $J_{E,F}^i$ are also time-independent.

It follows that e.g.

\beq \label{ExpIn} \langle \, \psi   \, |\, P_E^j(t) \, | \,   \phi    \, \rangle = \lim_{t \to -\infty}\langle \, \psi   \, |\, P_E^j(t) \, | \,   \phi    \, \rangle =\langle \, \psi   \, |\, P^j_{in}(E) \, | \,   \phi    \, \rangle \eeq
and analogously  for the angular momentum operators.\nl
\emph{Thus we have the important result that the nucleon matrix elements of $P^j_{E,F}$ and $J^j_{E,F}$ can be calculated by inserting a Fock expansion for the nucleon state and then evaluating the Fock state matrix elements of the ``in" field operators $P^j_{in}(E)$, $P^j_{in}(F)$, $J^j_{in}(E)$ and $J^j_{in}(F)$ respectively.}

\subsection{Interacting particles in gauge theories: canonical vs ``the rest"}

The objection of Ji, Chen et al and Wakamatsu to the use of the canonical operators is that they are not gauge invariant. We have suggested that this is not obviously important since the total canonical momentum and angular momentum operators are observables and their physical matrix elements  \emph{are} gauge invariant or BRST invariant (Sections~\ref{QED}B, \ref{QCD}A). That argument relied on the fact that an arbitrary physical state can be expressed as a superposition of eigenstates of \emph{total} momentum. \nl
Now from Eqs.~(\ref{Eq:inOp}-\ref{Eq:defPin}) and Eq.~(\ref{Eq:goodP}) it follows that, for $E$ (and analogously for $F$)
\beq \label{Eq:inCom} i[P^j_{in,\,can}(E),\, \phi_{in,\,E}(t,\bm{x})] = \partial^j  \phi_{in,\,E}(t,\bm{x}) \eeq

which implies that the Fock states, created from the vacuum by the action of the creation operators in $\phi_{in,\,E}(x), \, \phi_{in,\,F}(x)$,  are  eigenstates of  $P^j_{in,\,can}(E)$ and $P^j_{in,\,can}(F)$ respectively. This fact will be used in the next two sections in proving the gauge or BRST invariance of the Fock space matrix elements of $P^j_{in,\,can}(E)$ and $P^j_{in,\,can}(F)$.  An analogous statement holds for the angular momentum operators.

 \subsection{QED}
 Here particles $E$ and $F$ correspond to electrons and photons and the Fock states may be taken as superpositions of states with electrons having definite momentum $\bm{p_1}, \bm{p_2}---\bm{p_n}$    and transverse photons with momenta   $ \bm{k_1}, \bm{k_2}---\bm{k_m}$ . It is possible to show that these  eigenstates of $P^j_{in,\,can}(\textrm{electron})$ and $P^j_{in,\,can}(\textrm{photon})$ are physical states i.e.
 \beq \label{Eq:PhysIn}  B^{(+)}_{in}(x) |\, \bm{p_1}, \bm{p_2}---\bm{p_n}; \,  \bm{k_1}, \bm{k_2}---\bm{k_m} \rangle = 0. \eeq
 This follows from the asymptotic limit of the commutation relations given in \cite{Lautrup}, and the Greenberg-Robinson theorem \cite{Greenberg, Robinson}, which states that the commutators of asymptotic fields are c-numbers.

Since we are only concerned with physical matrix elements of the momentum operators we may, as a consequemce of Eq.~(\ref{Eq:PhysIn}) ignore the gauge-fixing terms and from now on utilise,
 \beq \label{Eq:Pelecsimpl} P^j_{in,\,can}(\textrm{electron})\equiv \int d^3x \left[\frac{i}{2}\, \bar{\psi}_{in}\gamma^0 \overleftrightarrow{\partial}^j \,\psi_{in}\right] \eeq
 and
 \beq \label{Eq:Pphotsimpl} P^j_{in,\,can}(\textrm{photon})\equiv \int d^3x \left[ - F^{0\beta}_{in}\partial^j A_{in,\,\beta} \right]. \eeq

 The proof that the Fock space matrix elements of these operators are gauge invariant requires that the matrix elements of $B_{in}(x)$ vanish between these states. This follows from Eq.~(\ref{Eq:PhysIn})  and thus the proof of the gauge invariance of the expectation values of $P^j_{in,\,can}(\textrm{electron})$ and $P^j_{in,\,can}(\textrm{photon})$ can be carried through in the same way as was done for the total momentum in Section~\ref{QED}B.

 Note that the simplified versions of the canonical momentum operators above generate the correct transformations on $\psi_{in}(x)$ and the spatial components $A_{in}^k(x)$, namely
 \beq \label{Eq:SimplTranscan} i[P^j_{in,\,can}(\textrm{electron}),\psi_{in}(t,\bm{x})]= \partial^j\psi_{in}(t,\bm{x}) \qquad i[P^j_{in,\,can}(\textrm{photon}), A_{in}^k(t,\bm{x})]=\partial^jA_{in}^k(t,\bm{x}). \eeq

 On the other hand one can show that the Bellinfante versions $P^j_{in,\,bel}(\textrm{electron})$ and $P^j_{in,\,bel}(\textrm{photon})$ do not generate the transformations Eq.~(\ref{Eq:SimplTranscan}). Thus the Bellinfante versions do \emph{not} satisfy our minimal requirement for identifying these operators as representing the momentum carried by the electrons and photons respectively. The same is true of the Chen et al and Wakamatsu momentum operators.

 The analysis of the  angular momentum operators is quite analogous and one concludes that the canonical operators are the ones that generate the correct rotations on the fields.

\subsection{QCD}

Similar results hold for QCD. The states with quarks having definite momentum $\bm{p_1}, \bm{p_2}---\bm{p_n}$    and transverse gluons having momenta   $ \bm{k_1}, \bm{k_2}---\bm{k_m}$ are  eigenstates of $P^j_{in,\,can}(\textrm{quark})$ and $P^j_{in,\,can}(\textrm{gluon})$ and are physical states i.e.
 \beq \label{Eq:PhysInQCD}  Q_{B} |\, \bm{p_1}, \bm{p_2}---\bm{p_n}; \,  \bm{k_1}, \bm{k_2}---\bm{k_m} \rangle = 0 \eeq
 This follows from the commutation relations for the asymptotic fields given in Section IV in \cite{Kugo:1977yx}.

 Since we are only concerned with the physical matrix elements of the momentum operators we may, as a consequence of the discussion following Eq.~(\ref{tbelQCD}), ignore the gauge-fixing and ghost terms and from now on utilise
 \beq \label{Pcanquarksimpl}  P^j_{in,\,can}(\textrm{quark})\equiv \int d^3x \left[\frac{i}{2}\, \bar{\psi}^l_{in}\gamma^0 \overleftrightarrow{\partial}^j \,\psi^l_{in}\right] \eeq
 and
 \beq \label{Pcangluonsimpl}   P^j_{in,\,can}(\textrm{gluon})\equiv \int d^3x \left[ - G^{0\beta}_{in,\,a}\partial^j A^a_{in,\,\beta} \right].                       \eeq
 These commute with $Q_{B}$ and are thus observables. Moreover these simplified versions of the canonical momentum operators generate the correct transformations on $\psi^l_{in}(x)$ and the spatial components  $A^k_{in,\,a}(x)$, namely
 \beq \label{SimpleTranscanQCD} i[P^j_{in,\,can}(\textrm{quark}),\psi^l_{in}t,\bm{x})]= \partial^j\psi^l_{in}(t,\bm{x}) \qquad i[P^j_{in,\,can}(\textrm{gluon}), A^k_{in,\,a}(t,\bm{x})]=\partial^jA^k_{in,\,a}(t,\bm{x}). \eeq

 On the other hand one can show that the Bellinfante versions $P^j_{in,\,bel}(\textrm{quark})$ and $P^j_{in,\,bel}(\textrm{gluon})$ do not generate the transformations Eq.~(\ref{SimpleTranscanQCD}). Thus the Bellinfante versions do \emph{not} satisfy our minimal requirement for identifying these operators as representing the momentum carried by the quarks and gluons. Similar remarks apply to the Chen et al and Wakamatsu operators.

 Similarly, one sees that the correct rotations of the fields are generated by the canonical versions of the angular momentum operators, which suggests that the Ji, Chen et al and Wakamatsu operators should not be regarded as representing the angular momentum of the quarks and gluons. Nonetheless, the expectation value of the Bellinfante operator $J_{z,\,bel}(\textrm{quark})$ used by Ji for the \emph{longitudinal} component of the quark angular momentum, which has the nice property that it can be measured in Deeply-virtual Compton Scattering reactions, does indeed represent the $Z$-component of the angular momentum carried by the quarks in a nucleon moving in the $Z$ direction, and therefore,  Ji's \emph{definition} of the \emph{orbital} angular momentum as the difference  $[ J_{z, \,bel}(\textrm{quark}) - \frac{1}{2}\Delta \Sigma_{\overline{MS}} \,]$, is fine as long as it is appreciated that this applies only to the components along the motion of the nucleon.

  \subsection{\label{longMom}The longitudinal component of the quark momentum and angular momentum }

  We have argued that the canonical versions of the momentum and angular momentum operators should be regarded as the physically meaningful ones. Yet
it is well known that $x_B$, Bjorken-$x$,  can be  interpreted as the fraction of the $Z$ component of the quark momentum  inside a nucleon, in an infinite momentum frame where the  nucleon is moving along the $OZ$ axis, and that this  corresponds, via the Operator Product Expansion, to the matrix element of the Bellinfante version of the momentum operators. At first sight this appears to contradict our assertion that it is the canonical version  that should be regarded as the physically meaningful  momentum operators. We shall here explain that there is, in fact, no  contradiction in the special case of the \emph{longitudinal}
components of the momentum and angular momentum. \nl
The gauge invariant expression for the unpolarized quark number density $q(x)$ is usually written as

\beq \label{q} q(x) =\frac{1}{2}\int \frac{dz^-}{2\pi}\, e^{i x P^+ z^-} \,\langle \, P \,| \,\bar{\psi}(-z^-/2)\, \gamma^+\, W \,\psi (z^-/2) \, | \, P \, \rangle|_{\,x>0} \eeq

where $|\, P \,\rangle $ corresponds to an unpolarized  proton moving along the $OZ$ axis i.e.
\beq P^\mu =(E,0,0,P), \eeq
and where
\beq \label{Wilson} W\equiv W[-z^-/2 \, , \, z^-/2]= \mathcal{P} \,\exp \{ig \int_{-z^-/2}^{z^-/2} \,dz'\,  A^+_a(z'n)\, t^a \} \eeq
 is the Wilson line operator, a matrix in colour space, and
where
\beq n=\frac{1}{\sqrt{2}}(1,0,0,-1). \eeq

We are using the standard definition of the $\pm $ components of a vector i.e.
\beq v^{\pm} = \frac{1}{\sqrt{2}}(v_0 + v_z).  \eeq

The expression for the antiquark density $\bar{q}(x)$ is  analogous to Eq.~(\ref{q}) but with $x<0$.
\nl

After some manipulation one finds that

\beqy \label{xq} x q(x) &=& \frac{i}{4P^+}\int \frac{dz^-}{2\pi}\, e^{i xP^+ z^-}\, \langle \, P \,| \,\{\bar{\psi}(z)[ -\overleftarrow{\partial}^+ -igA^+(z)] \}_{z=-z^-/2}\,\gamma^+ \, W  \, \psi(z^-/2)\nn \\
 &+& \bar{\psi}(-z^-/2)\, \gamma^+ \, W \,\{ [ \overrightarrow{\partial}^+\, - i g \, A^+(z)]\,\psi (z)\}_{z=z^-/2}  \, | \, P \, \rangle|_{\, x>0}. \eeqy
Integrating over $x$ one has
\beq \label{xq} \int_0^1 dx x\,[q(x) + \bar{q}(x) \,] = \frac{i}{4(P^+)^2}\langle \, P \,| \,\bar{\psi}(0) \, \gamma^+ \, \overleftrightarrow{D}^+ \,\psi (0) \, | \, P \, \rangle \eeq
 with
 \beq \overleftrightarrow{D}^+ = \overrightarrow{\partial}^+ - \overleftarrow{\partial}^+ -2ig A^+(0). \eeq

Now from Eq.~(\ref{Eq:tbelqG}) the quark part of $  t^{\mu\nu}_{bel}(qG)  $ is given by
\beq \label{tbelq} t^{\mu\nu}_{q, \, bel }(z)= \frac{i}{4}[\bar{\psi}(z) \gamma^\mu \overleftrightarrow{D}(z)^\nu \psi(z) + (\mu \leftrightarrow \nu )] - g^{\mu\nu}{\cal{L}}_{q}  \eeq
where ${\cal{L}}_{q}$ is the quark part of ${\cal{L}}_{qG}$ given in Eq.~(\ref{Eq:LqG}). \nl
Then, since $ g^{+ \, +} =0 $ we see that
\beq \label{Optbel} t^{+ \, +}_{q, \, bel}(0)\, = \frac{i}{2}\,\{\bar{\psi}(0)\, \gamma^+ \overleftrightarrow{D}^+\,\psi(0)\}   \eeq
so that
\beq \label{qmombel} \int_0^1 dx \, x\,[\,q(x) + \bar{q}(x) \,] = \frac{1}{2(P^+)^2} \langle \, P \, |\, t^{+ \, +}_{q, \, bel}(0)\,  \, | \, P \, \rangle. \eeq
Consider, now, the physical interpretation of the LHS of Eq.~(\ref{qmombel}) in the parton model. The parton model is not synonymous with QCD. It is a picture, a  manifestation, of QCD in the gauge $A^+ =0$ and it is in this gauge, and in an infinite momentum frame that $x$ can be interpreted as the momentum fraction carried by a quark in the nucleon. But since $ A^+= 0$ we have
\beq \label{Dd}  \,\overleftrightarrow{ D}^+ \,   = \overleftrightarrow{\partial}^+  \qquad  (\textrm{gauge} \, A^+=0) \eeq
so that  \emph{for these particular components} of the tensors there is no difference between the canonical and Bellinfante versions
\beq \label{tbtc} \,t^{+ \, +}_{q, \, can}(0)  = t^{+ \,+}_{q, \, bel}(0) \qquad  (\textrm{gauge} \, A^+=0).  \eeq
Hence the fraction of \emph{longitudinal} momentum carried by the quarks in an infinite momentum frame is given equally well by either the canonical or Belllinfante versions of the energy momentum tensor density.

Let us turn now to the question of the angular momentum and, in particular, to Ji's relation of the quark angular momentum to the second moment of certain generalized parton distributions (GPDs) measurable in Deeply Vitual Compton Scattering \cite{Ji:1996nm}. In the standard notation (see e.g. the review of Diehl \cite{Diehl:2003ny})

\beqy \label{GPD} &&\frac{1}{2}\int \frac{dz^-}{2\pi}\, e^{i x \bar{P}^+ z^-} \,\langle \, P' \,| \,\bar{\psi}(-z^-/2)\, \gamma^\mu \, W \,\psi (z^-/2) \, | \, P \, \rangle \nn \\
&=& \frac{1}{2 \bar{P}^+}\left\{ [\bar{u}(P')\gamma^\mu u(P)] H(x, \xi, t) + \left[\frac{i\Delta_\rho}{2M}\bar{u}(P') \sigma^{\mu\rho} u(P) \right] E(x,\xi, t) \right\} \eeqy
where
\beq \bar{P}=\frac{1}{2}(P + P')  \qquad \Delta = P'-P \qquad t= \Delta^2 \qquad \Delta^+= -2\xi\bar{P}^+ \eeq
and the spinors are normalized to $\bar{u}u=2M$.
Putting $P'=P$ i.e. $\Delta = t=\xi =0$ and comparing with Eq.~(\ref{q}) one sees that
\beq H(x,0,0) = q(x) \eeq
so that $xH(x,0,0)$ can be interpreted as the density in $x$-space of the quark longitudinal momentum. \nl
Now consider the general expression for the off-diagonal nucleon matrix element of $t^{\mu\nu}_{q, \, bel}(0)$. The connection between these matrix elements and the angular momentum involves divergent integrals, which have to be treated carefully using wave packets, as was done correctly for \emph{arbitrary} components of $\bm{J}$ for the first time by BLT \cite{Bakker:2004ib}, and for this reason we shall use their notation for the scalar functions that appear in the matrix element of $t^{\mu\nu}_{q, \, bel}(0)$. One has

\beqy \label{tmunu} \langle \, P',S' \, |\, t^{\mu\nu}_{q, \, bel}(0)\,  \, | \, P, S \, \rangle &=&[\bar{u}'\gamma^\mu u \, \bar{P}^\nu + (\mu \leftrightarrow \nu)]\mathbb{D}_{q, \, bel}(\Delta^2)/2 \nn \\
&& \hspace{-2cm}- \left[\frac{i\Delta_\rho}{2M}\, \bar{u}' \sigma^{\mu\rho} u \, \bar{P}^\nu + (\mu \leftrightarrow \nu) \right] [\mathbb{D}_{q, \, bel}(\Delta^2)/2 - \mathbb{S}_{q, \, bel}(\Delta^2)]\nn \\
 && \hspace{-4cm}+ \frac{\bar{u}' u}{2M}\left[\frac{1}{2}[\mathbb{G}_{q,\, bel}(\Delta^2)-\mathbb{H}_{q, \, bel}(\Delta^2)](\Delta^\mu \Delta^\nu - \Delta^2 g^{\mu\nu}) +M^2 \mathbb{R}_{q, \, bel}(\Delta^2)g^{\mu\nu}\right]
 \eeqy
 where
 \beq u\equiv u(P,S)  \qquad u'\equiv u(P',S'). \eeq
 Note that the term $M^2 \mathbb{R} g^{\mu\nu}$ is only allowed because we are dealing with  a non-conserved density. \nl
 Repeating for the GPDs the analysis which led to Eq.~(\ref{qmombel}) and bearing in mind Eq.~(\ref{GPD}) yields
 \beqy \label{intGPD}&& \frac{1}{2 \bar{P}^+}  \left\{ [\bar{u}'\gamma^+ u]\int dx x H(x, \xi, t) +  \left[\frac{i\Delta_\rho}{2M} \, \bar{u}' \sigma^{+ \rho} u \right] \int dx x E(x,\xi, t) \right\}\nn \\
 && \hspace{4cm}= \frac{1}{2(P^+)^2} \langle \, P',S' \, |\, t^{+ \, +}_{q, \, bel}(0)\,  \, | \, P,S \, \rangle. \eeqy
 From Eq.~(\ref{tmunu}), remembering that $g^{++}=0$ and that $\Delta^+= -2\xi \bar{P}^+$, one obtains
 \beqy \label{tplusplus}  \langle \, P',S' \, |\, t^{+ \, +}_{q, \, bel}(0)\,  \, | \, P,S \, \rangle &= &[\bar{u}'\gamma^+ u \, \bar{P}^+ ][\mathbb{D}_{q, \, bel}(\Delta^2) + \xi^2 ( \mathbb{G}_{q,\, bel}(\Delta^2)-\mathbb{H}_{q, \, bel}(\Delta^2)) ]\nn \\
  && \hspace{-4cm} + \left[\frac{i\Delta_\rho}{2M}\, \bar{u}' \sigma^{+\rho}u \, \bar{P}^+  \right] [  2 \,\mathbb{S}_{q, \, bel}(\Delta^2)- \mathbb{D}_{q, \, bel}(\Delta^2) - \xi^2 ( \mathbb{G}_{q,\, bel}(\Delta^2)-\mathbb{H}_{q, \, bel}(\Delta^2))]. \eeqy
  
  Comparing with Eq.~(\ref{intGPD}), taking the limit $\Delta \rightarrow 0 $ and writing $ \mathbb{D}_{q, \, bel} =  \mathbb{D}_{q, \, bel}(\Delta^2=0)$ etc,   one obtains
  
  \beq \label{Hsum} \int_{-1}^{1} dx x H(x,0,0) = \mathbb{D}_{q, \, bel} \eeq
  \beq \label{Esum} \int_{-1}^{1} dx x E(x,0,0)= (  2 \, \mathbb{S}_{q, \, bel} -\mathbb{D}_{q, \, bel}) \eeq
  and consequently
  \beq \label{HEsum} \int_{-1}^{1} dx x H(x,0,0) + \int_{-1}^{1} dx x E(x,0,0) = 2 \, \mathbb{S}_{q, \, bel}. \eeq
  Consider now the parton model interpretation of these expressions. Choosing the gauge $A^+=0$ we have, as before, $t^{+ \, +}_{q, \, can}(0)  = t^{+ \,+}_{q, \, bel}(0)$, so that in Eqs.(\ref{Hsum}, \ref{Esum}, \ref{HEsum}) we may put
  \beq  \mathbb{D}_{q, \, bel}= \mathbb{D}_{q, \, can} \equiv \mathbb{D}_{q} \qquad  \textrm{and} \qquad \mathbb{S}_{q, \, bel} =   \mathbb{S}_{q, \, can}\equiv \mathbb{S}_{q}. \eeq

For the case of a longitudinally polarized  nucleon moving at high speed in the $Z$ direction BLT \cite{Bakker:2004ib} proved that $\mathbb{S}$ measures the $Z$-component of $\bm{J}$. Hence Eq.~(\ref{HEsum}) can be written
\beq \label{Jzsum} \int_{-1}^{1} dx x [H(x,0,0) +E(x,0,0)]= 2 \, J_{z}(\textrm{quark}) \eeq
which is the relation first derived by Ji \cite{Ji:1996nm}. \nl
Note, however, that unlike the case of linear momentum, it is not obvious that $x\,[H(x,0,0) +E(x,0,0)]]$ can be interpreted as the $x$-space density of $J_z(\textrm{quark})$. Indeed, Burkardt and Hikmat \cite{Burkardt:2008ua} have shown, in a model,  that $J_z(\textrm{quark}; x)$ calculated directly from the nucleon wave function disagrees with $x\,[H(x,0,0) +E(x,0,0)]]$, whereas there is perfect agreement when integrated over $x$.

\subsection{Interacting particles: photons and gluons}

To a large extent the entire controversy concerning the assigning of angular momentum to quarks and gluons arose from the long established claim that one cannot split the angular momentum of a massless gauge particle into an orbital and spin part in a gauge-invariant way. The two standard expressions in the literature for the angular momentum for QED, the canonical and Bellinfante versions, are

\beqy \label{Jcan} \bm{J}_{can} &=& \int d^3x \, \psi^\dag \bm{\gamma} \gamma_5 \psi +\int d^3x \, \psi^\dag [\bm{x} \times (-i\bm{\nabla})] \psi \nn \\
&+& \int d^3x \,(\bm{E} \times \bm{A})  + \int d^3x \,E^i (\bm{x} \times \bm{\nabla}A^i) \nn \\
&=& \bm{S}_{can}(el) + \bm{L}_{can}(el) + \bm{S}_{can}(\gamma) + \bm{L}_{can}(\gamma)  \eeqy
and
\beqy \label{Jbel} \bm{J}_{bel} &=& \int d^3x \, \psi^\dag \bm{\gamma} \gamma_5 \psi +\int d^3x \, \psi^\dag [\bm{x} \times (-i\bm{D})] \psi \nn \\
&+& \int d^3x \,\bm{x} \times (\bm{E} \times \bm{B}) \nn \\
&=& \bm{S}_{bel}(el) + \bm{L}_{bel}(el) + \bm{J}_{bel}(\gamma) \eeqy
In $\bm{J}_{can}$ only the electron spin term is gauge invariant. In $\bm{J}_{bel}$ each of the three terms is gauge invariant, but the photon angular momentum is not split into a spin and orbital part. \nl
Insisting on  being able to split the photon angular momentum into a spin and orbital part, and on having each term gauge invariant,    Chen et al \cite{Chen:2008ag} arrived at the following form
\beqy \label{Jchen} \bm{J}_{chen} &=& \int d^3x \, \psi^\dag \bm{\gamma} \gamma_5 \psi +\int d^3x \, \psi^\dag [\bm{x} \times (-i\bm{D}_{pure})] \psi \nn \\
&+& \int d^3x \,(\bm{E} \times \bm{A}_{phys})  + \int d^3x \,E^i (\bm{x} \times \bm{\nabla}A^i_{phys}) \nn \\
&=& \bm{S}_{chen}(el) + \bm{L}_{chen}(el) + \bm{S}_{chen}(\gamma) + \bm{L}_{chen}(\gamma)  \eeqy
where $\bm{D}_{pure}= \bm{\nabla}-ie\bm{A}_{pure}$ and the fields $\bm{A}_{pure}$ and $\bm{A}_{phys}$ were explained in Eqs.~(\ref{Asplit}-\ref{Apure}) of the Introduction. \nl
Later, Wakamatsu \cite{Wakamatsu:2010qj} suggested a rearranged version of $\bm{J}_{chen}$, which retains a gauge-invariant split between the spin and orbital angular momentum of the photon

\beqy \label{Jwak} \bm{J}_{wak}  &=& \int d^3x \, \psi^\dag \bm{\gamma} \gamma_5 \psi +\int d^3x \, \psi^\dag [\bm{x} \times (-i\bm{D})] \psi \nn \\
&+& \int d^3x \,(\bm{E} \times \bm{A}_{phys})  + [\int d^3x \,E^i (\bm{x} \times \bm{\nabla}A^i_{phys})+ \int d^3x \, \psi^\dag (\bm{x}\times e\bm{A}_{phys})\psi ] \nn \\
&=& \bm{S}_{wak}(el) + \bm{L}_{wak}(el) + \bm{S}_{wak}(\gamma) + \bm{L}_{wak}(\gamma)  \eeqy
In this version the very last term $\int d^3x \, \psi^\dag (\bm{x}\times e\bm{A}_{phys})\psi$ has been shifted from Chen et al's electron orbital term to the photon's orbital angular momentum. We have already commented that one could do such a rearrangement in an infinite number of ways   by shifting some arbitrary fraction of this term.\nl
All of the above comments hold equally well for the case of QCD. \nl
As we have stressed, there is absolutely no need to have gauge-invariant operators so long as their physical matrix elements \emph{are} gauge invariant, as is the case for the canonical version of the electron spin, the electron orbital angular momentum and the photon's total angular momentum.  Moreover we have insisted that an angular momentum operator should generate rotations, at least in the restricted ``minimal" sense defined in Eq.~(\ref{Eq:goodM}). Only the canonical choice satisfies this requirement. We conclude, in agreement with the paper of Jaffe and Manohar \cite{Jaffe:1989jz}, that it is the terms in the canonical form $\bm{J}_{can}$ which should be interpreted as corresponding to the  angular momentum of the electron and photon respectively. Of course this leaves open the issue of splitting the photon angular momentum into spin and orbital parts. This we shall discuss in the next section.\nl

\section{\label{helicity} The spin of the photon and the gluon}

As has been  emphasized for more than half a century it is true that the \emph{canonical} photon  or gluon spin terms, as a whole, are not gauge invariant. This we regard as an inevitable feature of a gauge theory and it has not been the cause of any  problems in the description and calculation of physical processes involving photons, and more recently, gluons. However, the  projection of the spin terms onto the direction of the photon's or gluon's  momentum i.e. the photon and gluon helicity, \emph{is} gauge invariant and it is this quantity which can be measured  and, as we shall show,  is measured in deep inelastic scattering on atoms or nucleons respectively.
\subsection{ QED}
Consider the  expression for $\bm{S}_{can}(\gamma)$ in Eq.~(\ref{Jcan}), which can be written as

\beq \label{gamhel} S^k_{can}(\gamma) = \frac{1}{2}\, \epsilon_{kij}\,S^{i\,j}. \eeq

In accordance with the analogue of Eq.~(\ref{ExpIn}) we may study its expectation value between states of definite mass by replacing the fields by their ``in-field" versions  Thus we may utilize

\beq \label{Sjk}  S^{i\,j}_{in} = \int d^3x \, [F^{i\,0}_{in}(x)A^j_{in}(x) - F^{j\,0}_{in}(x)A^i_{in}(x)]. \eeq

We shall show  that the matrix element of the ``in-field" helicity operator

\beq \label{helOp} \mathcal{H}_{in}= \left[\, S^k_{can}(\gamma)P^k/|\bm{P}|\, \right]_{in}  \eeq

taken between arbitrary physical states of a photon is gauge invariant and then relate its matrix elements to the QED analogue of the polarized gluon density $\Delta g(x)$. \nl
Consider the action of $S^{i\,j}_{in}$ on a physical photon state of momentum $\bm{k}$ and polarization vector $\bm{\epsilon} (\bm{k} \,, l)$ corresponding to polarization along a transverse direction $l$ :

\beq \label{photstate} | \, \bm{k} \, , l \, \rangle = a^\dag (\bm{k} \, , l )\, |\, \textrm{vac} \,\rangle. \eeq

Provided the operators are  normal ordered we have,

\beq \label{Sjkcom} S^{i\,j}_{in} | \, \bm{k} \, , l \, \rangle  = [ S^{i\,j}_{in} \, , \, a^\dag (\bm{k} \, , l )]\,|\, \textrm{vac} \,\rangle.  \eeq

Then if $i$ and $j$ correspond to directions perpendicular to $\bm{k}$, expressing $a^\dag (\bm{k} \, , l )$ in terms of the fields as in Section 14.4 of \cite{Bjorken:1965dk}, and using  the equal time commutators (permitted because we are dealing with the ``in-field" momentum and angular momentum), gives

\beq \label{resSjkcom} [ S^{i\,j}_{in} \, , \, a^\dag (\bm{k} \, , l )] = i \{ \epsilon^i (\bm{k} \,, l)\, a^\dag (\bm{k} \, , j) -  \epsilon^j (\bm{k} \,, l)\, a^\dag (\bm{k} \, , i)\}. \eeq

Let us first check that acting on a helicity state, $\mathcal{H}_{in}$, as given by Eqs.~(\ref{helOp}, \ref{gamhel}), yields the correct result when using Eqs.~(\ref{Sjkcom}, \ref{resSjkcom}).      For simplicity take the $OZ$ axis along $\bm{k}$. Then

\beq \label{H} \mathcal{H}_{in}= \left[ \, S_{can,\, z}(\gamma)\, \right]_{in}= S^{1\,2}_{in} \eeq
and the helicity states are, for $\lambda = \pm 1 $,

\beq \label{helstate} |\, k \hat{\bm{z}}, \, \lambda \, \rangle = \frac{-\lambda}{\sqrt{2}}\, \{ |\, k \hat{\bm{z}}, \, 1 \, \rangle + i \,\lambda \,|\, k \hat{\bm{z}}, \, 2 \, \rangle \}. \eeq
 Using the fact that $\epsilon^j (k \hat{\bm{z}} \,, 1)  = \delta_{j1} $ , $\epsilon^j (k \hat{\bm{z}} \,, 2)  = \delta_{j2}$ one finds that indeed

 \beq \label{HelOK} \mathcal{H}_{in}\, |\, k \hat{\bm{z}}, \, \lambda \, \rangle = \lambda \, |\, k \hat{\bm{z}}, \, \lambda \, \rangle. \eeq

To show the gauge invariance of $\mathcal{H}_{in}$ we consider its action on a general physical photon state

\beq \label{genPhot} | \, \Phi \, \rangle = \int d^3 k \,\sum_{l\bot \bm{k}} \phi_l(\bm{k})  | \, \bm{k} \, , l \, \rangle \eeq
where the sum over $l$ refers to directions perpendicular to $\bm{k}$. Then
\beq \label{Hgen} \mathcal{H}_{in} \, | \, \Phi \, \rangle = \frac{1}{2}\int d^3 k \,\sum_{l\bot \bm{k}} \phi_l(\bm{k})\, \frac{k^r}{|\bm{k}|}\, \epsilon_{rij}\, S^{i \, j}_{in}| \, \bm{k} \, , l \, \rangle. \eeq
Since $i, j$ and $l$ refer to directions orthogonal to $\bm{k}$ we may use the results Eqs.~(\ref{Sjkcom}, \ref{resSjkcom}) to obtain
\beq \label{Hgenfin} \mathcal{H}_{in} \, | \, \Phi \, \rangle = \frac{i}{2}\int d^3 k \,\sum_{l\bot \bm{k}} \phi_l(\bm{k})\, \frac{k^r}{|\bm{k}|}\, \epsilon_{rij}\,
\{ \epsilon^i (\bm{k} \,, l)\, a^\dag (\bm{k} \, , j) -  \epsilon^j (\bm{k} \,, l)\, a^\dag (\bm{k} \, , i)\} \, |\, \textrm{vac} \, \rangle. \eeq
The creation operators in Eq.~(\ref{Hgenfin}) refer to polarization directions orthogonal to $\bm{k}$,  and are thus unaffected by gauge transformations. Hence the most general matrix element of $\mathcal{H}_{in}$ is gauge invariant.

For QED we now introduce $\Delta \gamma (x)$, the gauge-invariant analogue of $\Delta g(x)$, based on the expression for $\Delta g (x)$ given by Manohar          \cite{Manohar:1990kr} and used by Jaffe \footnote{Note that there is a typographical error in the expression for $\Delta g(x)$  in these papers:  $ \tilde{G}_{\alpha}^{\,\,\, +}(0)$ should be  $\tilde{G}^{+}_{\, \,\,\, \alpha}(0)$} \cite{Jaffe:1995an} i.e.
\beq \label{Delgamma} \Delta \gamma (x)= \frac{i}{4\pi x P^+}\int d\xi^- e^{-ix\xi^-P^+} \langle \bm{P}, \, S_L \,| F^{+\alpha}_{in}(\xi^-) I(\xi^-, 0) \tilde{F}^{+}_{\, \, \, \, \alpha, \, in}(0)\,| \bm{P}, \, S_L \,  \rangle + (x\rightarrow -x)\eeq
where $I(\xi^-, 0)$ is the Wilson line integral, and $| \bm{P}, \, S_L \,  \rangle $  is a longitudinally polarized, fast moving state. The axes are chosen so that $\bm{P}=(0,0,P)$ and we have used Eq.~(\ref{ExpIn}) to replace the fields by their ``in-field" versions.

Since the expression in Eq.~(\ref{Delgamma}) is gauge invariant we may evaluate it in the gauge $A^+=0$. Then following the argument in \cite{Jaffe:1995an} and integrating over $x$, we obtain

\beqy \label{delgam} \Delta \gamma &\equiv & \int dx \Delta \gamma(x) \nn \\
&=& \frac{1}{2P^+}\,\langle \bm{P}, \, S_L \,| F^{1 \,+}_{in}(0)A^2_{in}(0) -F^{2 \,+}_{in}(0)A^1_{in}(0)\,| \bm{P}, \, S_L \,  \rangle. \eeqy

Consideration of the possible tensorial structure for the matrix elements indicates that \emph{in leading twist}

\beq \label{LT} \langle \bm{P}, \, S_L \,| \, F^{i \,0}(0)| \bm{P}, \, S_L \,  \rangle = \langle \bm{P}, \, S_L \,| \, F^{i \,3}(0)| \bm{P}, \, S_L \,  \rangle \eeq
so that in leading twist

\beq \label{delgamLT} \Delta \gamma = \frac{1}{2E}\,\langle \bm{P}, \, S_L \,| F^{1 \,0}_{in}(0)A^2_{in}(0) -F^{2 \,0}_{in}(0)A^1(0)_{in}\,| \bm{P}, \, S_L \,  \rangle. \eeq

Now using Eq.~(\ref{Sjk}) one sees that
\beq \label{S12Mel} \langle \bm{P'}, \, S_L \,|\,S^{12}_{in}\,| \bm{P}, \, S_L \,  \rangle = (2\pi)^3\, \delta (\bm{P}' - \bm{P} ) \,\langle \bm{P'}, \, S_L \,| F^{1 \,0}_{in}(0)A^2_{in}(0) -F^{2 \,0}_{in}(0)A^1(0)_{in}\,| \bm{P}, \, S_L \,  \rangle. \eeq

Hence
\beq \label{delgamfin}  \Delta \gamma = \frac{\langle \bm{P'}, \, S_L \,|\,S^{12}_{in}\,| \bm{P}, \, S_L \,  \rangle}{2E (2\pi)^3\, \delta (\bm{P}' - \bm{P} )}. \eeq

But the denominator is just the norm of the state $| \bm{P}, \, S_L \,  \rangle$ so that $\Delta \gamma$ indeed measures the expectation value of the photon helicity operator.

\subsection{QCD}

Because we may use the ``in-fields" to study the matrix elements of the gluon helicity between arbitrary states of a nucleon, there is no essential difference  from the photon case.
The expression Eq.~(\ref{Sjk}) for $S^{ij}_{in}$ is simply altered by adding a colour label to the fields and summing over it. Similarly the expression for $\Delta g(x) $ and $ \Delta g $ are obtained from Eqs.~(\ref{Delgamma}) and (\ref{delgamLT}) by adding colour labels and summing over them. \nl
Thus $\Delta g $ indeed measures the expectation value of the gluon helicity in a nucleon.

\section{\label{Conc} Conclusions}

We have argued that there is no need to insist that the operators appearing in expressions for the momentum and angular momentum of the constituents of an interacting system should be gauge invariant, provided that the \emph{physical matrix elements} of these operators are gauge invariant. We have also suggested that the expressions given by  Chen et al and Wakamatsu for the momentum and angular momentum operators of quarks and gluons are somewhat arbitrary and do not satisfy the fundamental requirement that these operators should generate the relevant infinitesmal symmetry transformations specified in Eqs.~(\ref{Eq:goodP}, \ref{Eq:goodM}). Demanding that the conditions Eqs.~(\ref{Eq:goodP}, \ref{Eq:goodM}) be satisfied leads to the conclusion that  the \emph{canonical} expressions for the   momentum and angular momentum operators  are  the correct and physically meaningful ones. \nl
It is then an inescapable fact that the photon and gluon  angular momentum operators cannot, in general, be split in a gauge-invariant way into a spin and orbital part. However, as discussed in detail, the projection of the photon and gluon spin onto their direction of motion i.e. their helicity, is gauge-invariant and is measured in deep inelastic scattering on atoms or nucleons respectively.\nl
Although Ji's expressions for the quark and gluon angular momenta, which are the Bellinfante versions, do not conform to the above conditions and thus should not be considered as measuring arbitrary components of the quark and gluon momenta and angular momenta, nonetheless, it turns out that the expectation value of the Bellinfante operator $J_{z, \, bel}(\textrm{quark})$ used by Ji for the \emph{longitudinal} component of the quark angular momentum, which has the nice property that it can be measured in Deeply-virtual Compton Scattering reactions, does indeed represent the $Z$-component of the angular momentum carried by the quarks in a nucleon moving in the $Z$ direction, and therefore,  Ji's \emph{definition} of the \emph{orbital} angular momentum as the difference  $[ J_{z, \,bel}(\textrm{quark}) - \frac{1}{2}\Delta \Sigma_{\overline{MS}} \,]$, is fine as long as it is appreciated that this applies only to the components along the motion of the nucleon.

\section{Acknowledgements}
I am grateful to many colleagues, Professors Mauro Anselmino, Stan Brodsky, Matthias Burkardt, Marcus Diehl, Phillip Haegler, Roman Jackiw, Bob Jaffe, Taichiro Kugo, Benny Lautrup, Aneesh Manohar, Piet Mulders, Graham Shore and Masashi Wakamatsu, who were kind enough to respond to  a host of my queries, and to Professor Dave Websdale for helpful comments. I am also very grateful to Professors Feng Yuan, Jianwei Qiu and Kenji Fukushima for support and hospitality at the Berkeley Summer program on Nucleon Spin Physics (2009), the Brookhaven Summer Program on Nucleon Spin Physics (2010) and the Workshop on High Energy Strong Interactions at the Yukawa Institute for Theoretical Physics, Kyoto (2010), respectively.



\bibliography{Elliot_General}

\end{document}